\documentclass[twocolumn]{aastex63}

\usepackage{graphicx}
\usepackage[T1]{fontenc}
\usepackage{aecompl}
\usepackage[]{amsmath}

\usepackage{color}
\usepackage{float}
\usepackage{multirow}

\usepackage{xspace}
\usepackage{soul}
\usepackage{placeins}
\usepackage{atbegshi}
\newcommand{\handlethispage}{}
\newcommand{\discardpagesfromhere}{\let\handlethispage\AtBeginShipoutDiscard}
\newcommand{\keeppagesfromhere}{\let\handlethispage\relax}
\usepackage{lineno}

\newcommand{\msun}{\mathrm{M}_\odot}
\graphicspath{{figures/}}

\def\lsim{ \lower .75ex \hbox{$\sim$} \llap{\raise .27ex \hbox{$<$}} }

\submitjournal{ApJ}

\shorttitle{Satellite alignment}
\shortauthors{Yang et al.}

\graphicspath{{./}{figures/}}

\begin{document}

\title{The shape-velocity alignment of satellites forged by tidal locking and dynamical friction}

\correspondingauthor{Wenting Wang}
\email{wenting.wang@sjtu.edu.cn}

\author{Hao Yang}
\affiliation{State Key Laboratory of Dark Matter Physics, Tsung-Dao Lee Institute \& School of Physics and Astronomy, Shanghai Jiao Tong University, Shanghai 201210, China}
\affiliation{Key Laboratory for Particle Astrophysics and Cosmology (MOE)/Shanghai Key Laboratory for Particle Physics and Cosmology, Shanghai 200240,China}
\author[0000-0002-5762-7571]{Wenting Wang}
\affiliation{State Key Laboratory of Dark Matter Physics, School of Physics and Astronomy, Shanghai Jiao Tong University, Shanghai 200240, China}
\affiliation{Key Laboratory for Particle Astrophysics and Cosmology (MOE)/Shanghai Key Laboratory for Particle Physics and Cosmology, Shanghai 200240,China}
\author{Ting S. Li}
\affiliation{Department of Astronomy and Astrophysics, University of Toronto, 50 St. George Street, Toronto ON, M5S 3H4, Canada}
\author[0000-0003-2644-135X]{Sergey E. Koposov}
\affiliation{Institute for Astronomy, University of Edinburgh, Royal Observatory, Blackford Hill, Edinburgh EH9 3HJ, UK}
\affiliation{Institute of Astronomy, University of Cambridge, Madingley Road, Cambridge CB3 0HA, UK}
\author{Jiaxin Han}
\affiliation{Department of Astronomy, School of Physics and Astronomy, and Shanghai Key Laboratory for Particle Physics and Cosmology, Shanghai Jiao Tong University, Shanghai 200240, People's Republic of China}
\affiliation{State Key Laboratory of Dark Matter Physics, School of Physics and Astronomy,Shanghai Jiao Tong University, Shanghai 200240, China}
\author[0000-0003-0303-4188]{Feihong He}
\affiliation{Department of Astronomy, School of Physics and Astronomy, and Shanghai Key Laboratory for Particle Physics and Cosmology, Shanghai Jiao Tong University, Shanghai 200240, People's Republic of China}
\affiliation{State Key Laboratory of Dark Matter Physics, School of Physics and Astronomy,Shanghai Jiao Tong University, Shanghai 200240, China}
\affiliation{Kavli Institute for Astronomy and Astrophysics, Peking University, Beijing, 100871, China}
\author[0000-0001-7890-4964]{Zhaozhou Li}
\affiliation{School of Astronomy and Space Science, Nanjing University, Nanjing 210093, China}
\affiliation{Key Laboratory of Modern Astronomy and Astrophysics, Nanjing University, Ministry of Education, Nanjing 210093, China}
\author{Zhongxu Zhai}
\affiliation{Department of Astronomy, School of Physics and Astronomy, and Shanghai Key Laboratory for Particle Physics and Cosmology, Shanghai Jiao Tong University, Shanghai 200240, People's Republic of China}
\affiliation{State Key Laboratory of Dark Matter Physics, School of Physics and Astronomy,Shanghai Jiao Tong University, Shanghai 200240, China}
\author{Binbin Gao}
\affiliation{Department of Astronomy, School of Physics and Astronomy, and Shanghai Key Laboratory for Particle Physics and Cosmology, Shanghai Jiao Tong University, Shanghai 200240, People's Republic of China}
\affiliation{State Key Laboratory of Dark Matter Physics, School of Physics and Astronomy,Shanghai Jiao Tong University, Shanghai 200240, China}
\author{Carles G. Palau}
\affiliation{Department of Astronomy, School of Physics and Astronomy, and Shanghai Key Laboratory for Particle Physics and Cosmology, Shanghai Jiao Tong University, Shanghai 200240, People's Republic of China}
\affiliation{State Key Laboratory of Dark Matter Physics, School of Physics and Astronomy,Shanghai Jiao Tong University, Shanghai 200240, China}
\author{Zhenlin Tan}
\affiliation{Department of Astronomy, School of Physics and Astronomy, and Shanghai Key Laboratory for Particle Physics and Cosmology, Shanghai Jiao Tong University, Shanghai 200240, People's Republic of China}
\affiliation{State Key Laboratory of Dark Matter Physics, School of Physics and Astronomy,Shanghai Jiao Tong University, Shanghai 200240, China}

\begin{abstract}

Utilizing the TNG50 simulation, we study two types of alignments for satellites/subhalos: 1) the alignment of their major axes with the galactocentric radial directions (radial alignment), and 2) with the motion directions (orbital alignment). We find that radial alignment is substantially stronger than orbital alignment, with both signals being consistently stronger for subhalos than for satellites. Interestingly, inward- and outward-moving satellites/subhalos show contrasting orbital alignment behaviors, which can be understood in terms of their radial alignment, orbit decay due to dynamical friction and the effect of tidal stripping. The orbital alignment is stronger in more massive halos. In the end, we explore the orbital alignment measured by a mock observer, and find that the observed alignment for MW satellites is due to projection effects, as the major axes of satellites lie within their orbital planes, approximately coplanar with the observer.
\end{abstract}


\section{Introduction}
\label{sec:intro}

According to the standard Lambda cold dark matter ($\Lambda$CDM) cosmological model, galaxies form through gas condensation within dark matter halos \citep[e.g.][]{White1978,White1991}. Smaller halos merge into larger halos and become subhalos, and the galaxies formed in these smaller halos become satellite galaxies orbiting around the central galaxies in the massive hosts. The complex mechanism behind galaxy formation and evolution is encoded in numerous observable properties of galaxies \citep[for a more complete introduction, see e.g.][]{Mo2010}. Galaxy intrinsic alignments are one of such properties that reveal the spatial distribution of galaxies, connections to each other and connections to the large cosmic web of dark matter. They are an important source of systematic errors of weak lensing, but carry important information about the assembly history of galaxies and their host halos \citep[see][for reviews]{Joachimi2015,Kiessling2015,2015SSRv..193..139K}, and can serve as important cosmological probes \citep[e.g.][]{2013JCAP...12..029C,2018JCAP...08..014K,2023ApJ...945L..30O,2023NatAs...7.1259X}.

In this paper we are interested in central-satellite alignments within dark matter halos. On such small scales, there are different types of galaxy intrinsic alignment signals. In \cite{Mandelbaum2005}, the correlation between the spatial distribution of satellite galaxies and the orientation of their central galaxies/host halos is referred to as the density-shape alignment. The correlation between the major axes of satellite galaxies and those of central galaxies is referred to as the shape-shape alignment.

After some early debates on the density-shape alignment \citep[e.g.][]{Holmberg1969,Valtonen1978,Zaritsky1997}, using large samples of galaxy surveys and numerical simulations, it has been widely accepted that satellite galaxies tend to distribute more around the major axes of central galaxies, and this distribution depends on properties of both central and satellite galaxies \citep[e.g.][]{Brainerd2005,Yang2006,Agustsson2006b,Kang2007,Faltenbacher2007,Faltenbacher2008,Agustsson2010,Dong2014,Velliscig2015,Wang2018,Tang2020,Tenneti2021,Rodriguez2022,Lan2024}. For example, the spatial distribution of red satellites around red central galaxies is the most anisotropic, while that of satellites around blue central galaxies is nearly isotropic. The stronger anisotropy in the spatial distribution of red satellites, which reside in more massive subhalos, is due to the preferential accretion along the major axes of host halos. Moreover, recent observations and simulations have found that more satellites tend to be located at one side of their central galaxies \citep[e.g.][]{Libeskind2016,Pawlowski2017,Brainerd2020,Wang2021}. This "lopsided satellite distribution" can be identified around both galaxy pairs and isolated galaxies. It has also been suggested that the lopsidedness originates from the anisotropic satellite accretion with respect to the large scale structure \citep{Gong2019,Liu2024}, such as along the filament direction \citep[e.g.][]{Kang2015,Libeskind2015,Shao2018}.

The shape-shape alignment signal between satellites and their central galaxies is relatively weak. Most studies only found weak but still significant alignment signals in the innermost regions of clusters, but with an absence in outer regions \citep[e.g.][]{Plionis2003,Agustsson2006a,Mandelbaum2006,Faltenbacher2007,Faltenbacher2008,Torlina2007,Lan2024}.The shape-shape alignment is believed to result from the long-term tidal effects on satellites exerted by the anisotropic matter distribution of their host dark matter halos \citep[e.g.][]{Lee2005,Lee2008}. However, more evidence and investigations on this alignment signal are needed.

The correlation of the major axes of satellite galaxies and their radial directions towards the center of the host halo is called the radial alignment \citep[e.g.][]{Faltenbacher2007}. The radial alignment has been observed by many surveys and can also be replicated by numerical simulations \citep[e.g.][]{Hawley1975,Pereira2005,Agustsson2006a,Faltenbacher2007,Knebe2008,Schneider2013,Rong2015,Wang2019,Knebe2020,Tenneti2021,Lan2024}. Note that there are measurements with no alignment signals \citep[e.g.][]{Chisari2014,Sifon2015}, which is because they include fainter blue satellites diluting the signal of luminous red satellites \citep{Singh2015}. The strength of the signal also shows some dependence on the distance to the center of the halo, but is independent of the mass of satellites/subhalos. This type of alignment is assumed to be caused by the tidal torque of host halos \citep{Ciotti1994,Usami1997}, and is correlated with orbital phases of satellites/subhalos \citep{Pereira2008}.

In addition to the density-shape, shape-shape and radial alignments for satellite galaxies mentioned above, various studies also exist in studying the alignment of galaxies and their host dark matter halos, through the correlations of galaxy ellipticities or gravitational shear and galaxy ellipticity correlations \citep[e.g.][]{Okumura2009,Okumura2009b,2023ApJ...957...45X,2023ApJ...954....2X}. There are also attempts directly looking at the spin alignments between different types of objects within galaxy groups/clusters, including the spin of central and satellite galaxies, and the spin of gas \citep[e.g.][]{2024MNRAS.527.7028W,2025ApJ...983..100W,2025ApJ...986...85W,2025ApJ...992L..17W}.

Compared with more distant galaxy systems, satellites in our Milky Way (MW) are close enough to have their systemic velocities precisely determined, especially their proper motions. Based on \textit{Gaia} EDR3 astrometry \citep{Gaia2021}, \cite{Pace2022} measured proper motions of 52 dwarf spheroidal (dSph) satellite galaxies in our MW. They found a correlation between the orbital motion direction and the orientation of these satellites, which, at face value, is different from the other types of galaxy intrinsic alignments summarized above. Specifically, the major axes of satellites with large ellipticities (i.e. ellipticities larger than 0.4) tend to be aligned with their orbital motions. \cite{Pace2022} argue that this alignment may be a signature of tidal disruption, but they do not find similar trend for satellites with small pericenters which have experienced strong tidal stripping. Alternatively, it could be a projection of the radial alignment as expected from tidal torque mentioned above. However, they lack a complete sample of MW satellites to confirm this explanation. In this work we will explore this orbital alignment in the state-of-the-art hydrodynamical simulations.

This paper is organized as follows. We first introduce our sample of satellites/subhalos selected from simulation data in Section \ref{sec:data}. We describe the methods to determine the orientation of satellites/subhalos and how we quantify the alignment signal in Section \ref{sec:method}. Our results are presented in Section \ref{sec:results}, including alignments in both 3-dimensional and 2-dimensional space. We conclude in Section \ref{sec:concl}.

\section{Data}
\label{sec:data}

\subsection{The IllustrisTNG simulation}

We utilize the IllustrisTNG suites of simulations in this work. Based on the moving-mesh code \citep[\textsc{arepo};][]{Springel2010}, IllustrisTNG simulations incorporate sophisticated baryonic processes such as star formation and evolution, black hole accretion and feedback, chemical enrichment and gas recycling \citep{Weinberger2017,Marinacci2018,Naiman2018,Nelson2018,Pillepich2018,Springel2018,Nelson2019}. The Planck 2015 $\Lambda$CDM cosmological model with $\Omega_\mathrm{m}=0.3089$, $\Omega_\Lambda=0.6911$, $\Omega_\mathrm{b}=0.0486$, $\sigma_8=0.8159$, $n_s=0.9667$, and $h=0.6774$ \citep{Planck2015} is adopted, and the initial condition is set up at $z=127$. The simulations save 100 snapshots spanning the redshift range from $z=20$ to $z=0$.

The TNG50-1 simulation is used in our analysis, which has the highest resolution in its suite, compared with TNG50-2 and TNG50-3. TNG50-1 traces the joint evolution of 2,160$^3$ dark matter particles and $\sim$2,160$^3$ gas cells. The mass of each dark matter particle is $3.1\times10^5 \msun$/h, while the initial mass of each baryonic particle is $5.7\times10^4 \msun$/h. Hereafter we refer to TNG50-1 as TNG50 for short.

\subsection{Our sample of satellite galaxies}

We study satellite galaxies and their subhalos from TNG50 at redshift $z=0$. Note that throughout this paper, subhalos are defined through the bound dark matter particles, while satellite galaxies are the stellar components in the center of these subhalos. We first select satellite galaxies within the virial radius\footnote{The virial radius $R_{200,\mathrm{host}}$ of the host dark matter halo is defined as the radius within which the total matter density is 200 times the critical density of the universe.}, $R_{200,\mathrm{host}}$, of their host halos, and also discard those at small galactocentric radii ($r<0.2 R_{200,\mathrm{host}}$). To ensure reliable measurements of the shapes of these satellites and their subhalos, we restrict our analysis to those containing more than 100 star particles and also more than 100 dark matter particles in subhalos. After these selections, there are 4798 satellites/subhalos in this entire sample. Note for the entire sample, we do not include any additional selection on the stellar or subhalo mass for satellites, but the requirement of at least 100 star particles selects satellites that are on average more massive than $\sim8.5 \times 10^6 M_\odot$. To study the dependence of possible signals on the host halo mass, we bin satellites/subhalos according to the virial mass\footnote{The virial mass $M_{200},\mathrm{host}$ of the host dark matter halo is defined as the total mass enclosed within the virial radius $R_{200,\mathrm{host}}$.} $M_{200,\mathrm{host}}$ of their host halos, such that different bins contain similar amounts of satellites/subhalos, which are provided in Table \ref{tab:num per bin}. In the following we use $M_{\mathrm{host}}$ and $R_{\mathrm{host}}$ to denote the virial mass and the virial radius of host halos for short.

To compare with observations of Milky Way (MW) satellites in \cite{Pace2022}, we also use the TNG50 Milky Way $+$ Andromeda-like (MW/M31-like) sample. Specifically, we use the catalog of MW/M31-like central galaxies presented in \cite{Pillepich2024} and the catalog of their satellite galaxies presented in \cite{Engler2021} and \cite{Engler2023}, where MW/M31-like centrals are galaxies with stellar masses in the range of $M_\star=10^{10.5-11.2}M_\odot$ in relative isolation at $z = 0$, and their satellites are galaxies with $M_\star \ge 5 \times 10^6 M_\odot$ within $300$kpc of these MW/M31-like centrals. In addition, we also require these satellites in the MW/M31-like systems should have at least 100 star particles and 100 dark matter particles, and we remove satellites at galactocentric radii smaller than $0.2R_\mathrm{host}$, like the entire sample. There are 979 satellites/subhalos in this MW/M31-like sample.

\begin{table*}
\centering
\begin{tabular*}{0.8\textwidth}{@{\extracolsep{\fill}}cccccc}
\hline
\multirow{3}{*}{$\log_{10}(M_\mathrm{host}/M_\odot)$}
& \multicolumn{5}{c}{Number of satellites/subhalos in the entire sample} \\ \cline{2-6}
& \multirow{2}{*}{Total}
& $r<\frac{1}{2}R_\mathrm{host}$
& $r<\frac{1}{2}R_\mathrm{host}$
& $r>\frac{1}{2}R_\mathrm{host}$
& $r>\frac{1}{2}R_\mathrm{host}$ \\
& & move in & move out & move in & move out \\ 
\hline
10.0--12.0 & 1061 & 273 & 208 & 381 & 199 \\
12.0--12.9 & 1297 & 287 & 243 & 417 & 350 \\
12.9--13.6 & 1479 & 298 & 275 & 502 & 404 \\
13.6--14.5 & 961  & 202 & 189 & 313 & 257 \\
\hline
\end{tabular*}
\caption{Numbers of satellites/subhalos in different bins of the entire sample. The first and second columns show the host halo mass bins and the total numbers of satellites/subhalos in each bin. The four columns on the right show the numbers of satellites/subhalos with different orbital phases in each host halo mass bin, where they are classified into four groups according to their galactocentric radii $r$ and directions of their radial velocities.}
\label{tab:num per bin}
\end{table*}

\section{Method}
\label{sec:method}

To determine the orientation of a system (a satellite or a subhalo), we first compute the moment of inertia (MoI) tensor $I_{ij}$:
\begin{equation}
I_{ij} \equiv \sum_{k=1} m_k \, r_{k,i} \, r_{k,j},
\label{eq:moi}
\end{equation}
where $m_k$ is the mass of particle $k$, and $r_{k,i}$ is the $i$-th ($i=1,2,3$) component of the position vector $\boldsymbol{r}_k$ for particle $k$, with respect to the center of the system. The sum is over all particles (star particles for satellite galaxies, and dark matter particles for subhalos) bound to this system. In our analysis, we assume that all particles in a system have equal mass (i.e., $m_k=1$). This assumption is exact for dark matter particles in subhalos, but is an approximation for star particles in satellites. We have verified that the results for satellites remain nearly the same when true masses of star particles are used.

The shape and orientation of the system are then determined by the eigenvalues $\lambda_i$ ($i=1,2,3$ and $\lambda_1 \ge \lambda_2 \ge \lambda_3$) and the eigenvectors $\boldsymbol{e}_i$ of $I_{ij}$. The rescaled lengths of major, intermediate and minor axes are $a=\sqrt{\lambda_1/N}$, $b=\sqrt{\lambda_2/N}$ and $c=\sqrt{\lambda_3/N}$, where $N$ is the total number of star/dark matter particles in the satellite/subhalo. The corresponding eigenvectors are $\boldsymbol{e}_1$, $\boldsymbol{e}_2$ and $\boldsymbol{e}_3$, respectively. The ellipticity $e$ of this system is defined as $e=1-c/a$. We will use the direction of the major axis (i.e., the corresponding eigenvector $\boldsymbol{e}_1$) to represent the orientation of this system.

The alignment between the satellite/subhalo's major axis and the direction of its velocity/position with respect to the host halo will be explored in this work. The velocities and positions of satellites are the same as those of their subhalos, which are provided in outputs of TNG50. We quantify the alignment between any two directions along vectors $\boldsymbol{v}_1$ and $\boldsymbol{v}_2$ through an alignment angle $\theta$ ($0^\circ \le \theta \le 90^\circ$), which is defined as:

\begin{equation}
    \theta = \arccos{\left( \left| \boldsymbol{\hat{v}_1}\cdot\boldsymbol{\hat{v}_2}  \right| \right)},
    \label{eqn:theta}
\end{equation}
where $\boldsymbol{\hat{v}}_1$ and $\boldsymbol{\hat{v}}_2$ are unit vectors of $\boldsymbol{v}_1$ and $\boldsymbol{v}_2$, respectively. By definition, $\theta \sim 0^\circ$ indicates a perfect alignment, whereas $\theta \sim 90^\circ$ indicates a perpendicular misalignment. Statistically, we will study the distribution of the alignment angle $\theta$ for our sample of satellites/subhalos. For two randomly oriented vectors in 3-dimensional space,  $\cos{(\theta)}$ is uniformly distributed because the relative orientations of the two vectors should sample a sphere uniformly, with the area of the surface element $\propto \sin{(\theta)} \mathrm{d}\theta$. Whereas in 2-dimensional space, the angle $\theta$ itself is uniformly distributed because the relative orientations should sample a circle uniformly, the length of whose line element $\propto \mathrm{d}\theta$. The deviation from the uniform distribution of the corresponding quantity in each case is regarded as alignment or misalignment signals, depending on whether the distribution is more biased to $\theta\sim0^\circ$ or 90$^\circ$.

The alignment angle defined in Equation~\ref{eqn:theta} ranges from 0 to 90$^\circ$, as we do not distinguish the sign of $\boldsymbol{\hat{v}_1}\cdot\boldsymbol{\hat{v}_2}$. Throughout this paper, we will first show the measured radial and orbital alignments based on the definition in Equation~\ref{eqn:theta} above, which is also more closely connected to real observations. Then in order to draw more intuitive conclusions from our measurements, we also investigate another alignment angle defined as

\begin{equation}
    \alpha = \arccos{\left(\boldsymbol{\hat{v}_1}\cdot\boldsymbol{\hat{v}_2} \right)}.
    \label{eqn:alpha}
\end{equation}
The angle $\alpha$ ranges from 0 to 180$^\circ$. Here for the major axes of satellites/subhalos, we define their positive directions as pointing to the central galaxies of their host halos.

\section{Results}
\label{sec:results}

\subsection{Alignments in 3-dimensional space: radial alignment}
\label{subsec: 3d radial alignment}

\begin{figure}[htbp!]
    \centering
    \includegraphics[width=0.49\textwidth]{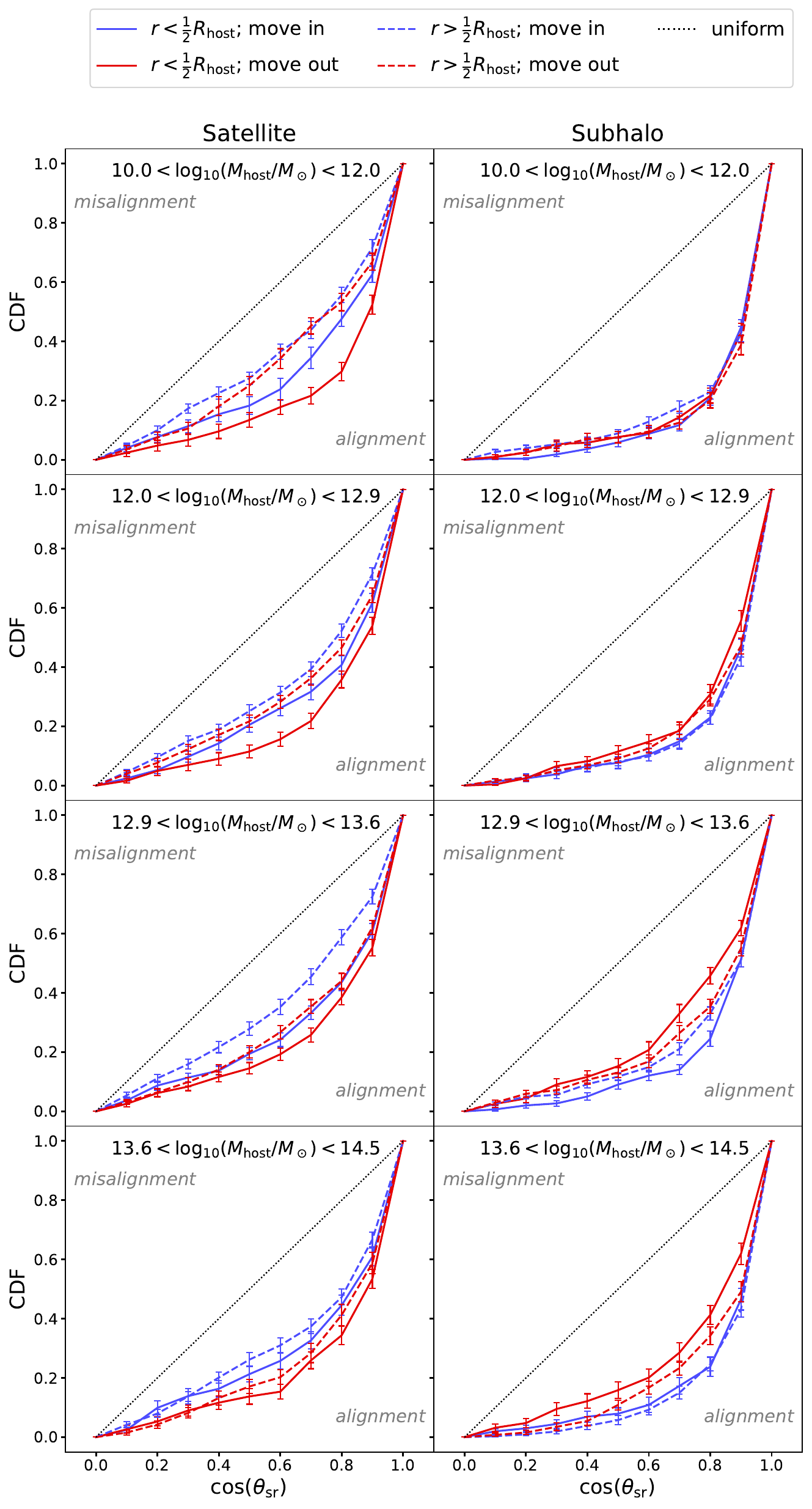}
    \caption{The cumulative distribution function (CDF) of the cosine of the radial alignment angle $\theta_\mathrm{sr}$ between the major axes of the systems and their galactocentric radial directions. Each row corresponds to a given mass bin in $\log_{10}(M_\mathrm{host}/M_\odot)$ of their host halos. The left column shows the results of satellites, and the right column shows the results of their subhalos. The blue (red) curves refer to systems moving inwards (outwards). The solid (dashed) curves refer to systems at the galactocentric radii $r$ smaller (larger) than the half of the virial radii $R_\mathrm{host}$ of their host halos. The black dotted line in each panel denotes the CDF of a uniform distribution, and curves above (below) it correspond to misalignment (alignment) signals. The error bars represent the 1$\sigma$ scatters of 50 bootstrapped subsamples of the systems in the same bin.}
    \label{fig:thetasr}
\end{figure}

We first check the alignments between the major axes of satellites/subhalos and their galactocentric radial directions (radial alignment) in the entire sample, which is expected to exist due to tidal torque \citep[e.g.][]{Ciotti1994,Usami1997,Pereira2008}, and we will show that it is related to the orbital alignment. The galactocentric radial direction is defined as the line connecting the satellite/subhalo and its corresponding central galaxy. Here we adopt the symbol $\theta_\mathrm{sr}$ to quantify the angle between the two directions, with the lower index sr referring to shape (major axis)-radius (radial direction). Figure \ref{fig:thetasr} shows the cumulative distribution function (CDF) for the cosine of $\theta_\mathrm{sr}$. Note that for random orientations in 3-dimensional space, it is the cosine of the alignment angle that follows a uniform distribution, which is denoted by the diagonal black dotted line to show the cumulative distribution in each panel. We also present results for satellites/subhalos for the MW/M31-like sample in Appendix \ref{appsec:3d alignment mw sample}, which show consistent trends with Figure \ref{fig:thetasr}.

Figure \ref{fig:thetasr} shows that both satellite galaxies and their subhalos show strong radial alignments, and the alignments of subhalos are stronger than those of satellites. Our results are consistent with those in previous works \citep[e.g.][]{Pereira2008,Knebe2020,Lan2024}. The radial alignment is due to the tidal torque exerted by the host halos, which turns the systems' major axes towards their galactocentric radial directions. Moreover, subhalos are more extended than satellites, and in turn are more sensitive to tidal torque \citep{Pereira2008}. Therefore, subhalos feature stronger alignment signals. However, there is relatively little dependence on the host halo mass (different rows), likely indicating the self-similar or scale free nature of gravitational forces. 

Notably, some studies \citep[e.g.][]{Pereira2008,Knebe2020} used the \textit{reduced} MoI tensor\footnote{The \textit{reduced} MoI tensor $\tilde{I}_{ij}$ is defined as $\tilde{I}_{ij} \equiv \sum_{k=1} m_k \, r_{k,i} \, r_{k,j}/ r_k^2$, where $r_k$ is the distance for particle $k$ to the center of the system, and other symbols have the same meanings as in Equation \ref{eq:moi}. $\tilde{I}_{ij}$ features an additional $1/r^2$ weighting for each particle compared with the MoI tensor $I_{ij}$ defined in Equation \ref{eq:moi}.} to determine the shape and orientation of a system, which assigns particles in inner regions more weights when calculating the MoI in Equation \ref{eq:moi}. We have checked that our results are consistent using either the original MoI or the reduced MoI tensor, but the results computed from the \textit{reduced} MoI tensor is less significant, which is expected, because the inner particles with larger weights are affected less by tides. Thus, we show results from the original MoI tensor with more significant signals in the main text, and present example results from the reduced MoI tensor in Appendix \ref{appsec:red moi}.

To investigate possible correlations between the radial alignment and orbital phases of these systems, in each panel of Figure \ref{fig:thetasr} we divide satellites/subhalos into four groups according to their galactocentric radii and directions of their radial velocities, and the numbers of them in each group are shown in Table \ref{tab:num per bin}. Note that systems with positive radial velocities are defined as moving outwards, while those with negative radial velocities are defined as moving inwards. The strength in alignment signals is different for distinct orbital phases, especially for satellites. In the left column of Figure \ref{fig:thetasr}, satellites in inner regions of their hosts show stronger alignment signals (solid curves) than those in outer regions (dashed curves). Moreover, satellites moving outwards have stronger alignment signals than those moving inwards. However, the situation is different for subhalos in the right column of Figure \ref{fig:thetasr}. Subhalos moving inwards (blue curves) have stronger signals than those moving outwards (red curves). The dependence of the signal on the orbital phase for satellites is different from that for subhalos. In particular, the two top panels on the right column reveal little dependence on the orbital phase, while the two bottom rows show opposite dependence on the orbital phase for satellites and subhalos. This is likely because more extended subhalos respond more quickly to tidal torque, which results in the alignment signal before passing the pericenter (moving inwards). On the other hand, it takes longer time for satellites embedded in subhalos to be influenced, which explains why the stronger signal happens when the satellite passes the pericenter (moving outwards). This also explains why satellites in inner regions of their hosts, where tidal torque is stronger, show more prominent signals than those in outer regions.

\subsection{Alignments in 3-dimensional space: orbital alignment}
\label{subsec: 3d orbital alignment}

\begin{figure}[htbp!]
    \centering
    \includegraphics[width=0.49\textwidth]{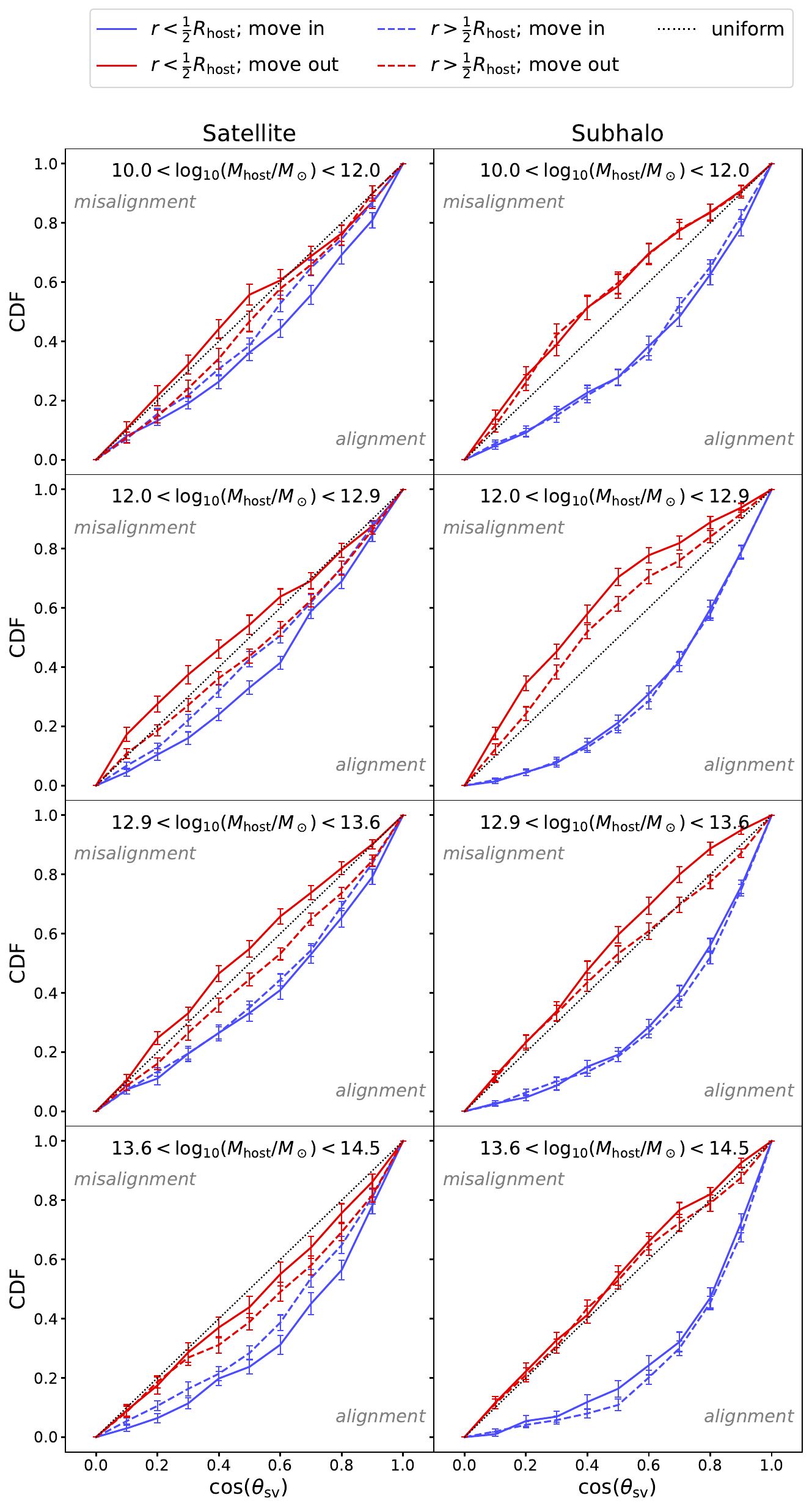}
    \caption{Similar to Figure \ref{fig:thetasr}, but of the cosine of the orbital alignment angle $\theta_\mathrm{sv}$ between the major axes of the systems and their motion directions.}
    \label{fig:thetasv}
\end{figure}

We then move on to study the alignments between the major axes of satellites/subhalos and their motion directions (orbital alignment) in the entire sample. Here we adopt the symbol $\theta_\mathrm{sv}$ to denote the angle between the two directions, with the lower index sv referring to shape (major axis)-velocity (motion direction). Figure~\ref{fig:thetasv} shows the cumulative distribution function for the cosine of $\theta_\mathrm{sv}$. Similar results for satellites/subhalos in the MW/M31-like sample are shown in Appendix \ref{appsec:3d alignment mw sample}. Generally, the signals in Figure \ref{fig:thetasv} are weaker than those in Figure \ref{fig:thetasr}, with less deviations from the uniform distribution (black dotted lines). Comparing the two columns in Figure \ref{fig:thetasv}, we find signals are still stronger in subhalos than in satellites, indicating the alignment is also more prominent for particles in more outskirts of the systems.

We see that satellites in inner regions of their hosts exhibit slightly stronger alignment signals than those in outer regions, with the blue solid curves a bit lower than the blue dashed curves. Interestingly, satellites/subhalos with different orbital phases tend to show totally opposite orbital alignment signals, and the dependence on the orbital phase is significantly more prominent than those in Figure~\ref{fig:thetasr}. For systems moving inwards (blue curves), their major axes tend to align with their motion directions, while for those moving outwards (red curves), their major axes tend to be instead more misaligned with their motion directions, which we will back discussing the possible causes in Section~\ref{subsec:cause} below.

Moreover, we find that with the increase of the host halo mass ($M_{\mathrm{host}}$) in different rows, the alignment of systems moving inwards (blue curves) become stronger, while those moving outwards (red curves) become less misaligned with their motion directions. Note that we do not see significant dependence of the radial alignment signal on the host halo mass (Figure~\ref{fig:thetasr}). We will revisit this point in Section~\ref{subsec:cause} below.

\subsection{Reasons behind the satellite/subhalo orbital alignment}
\label{subsec:cause}

\begin{figure}[htbp!]
    \centering
    \includegraphics[width=0.49\textwidth]{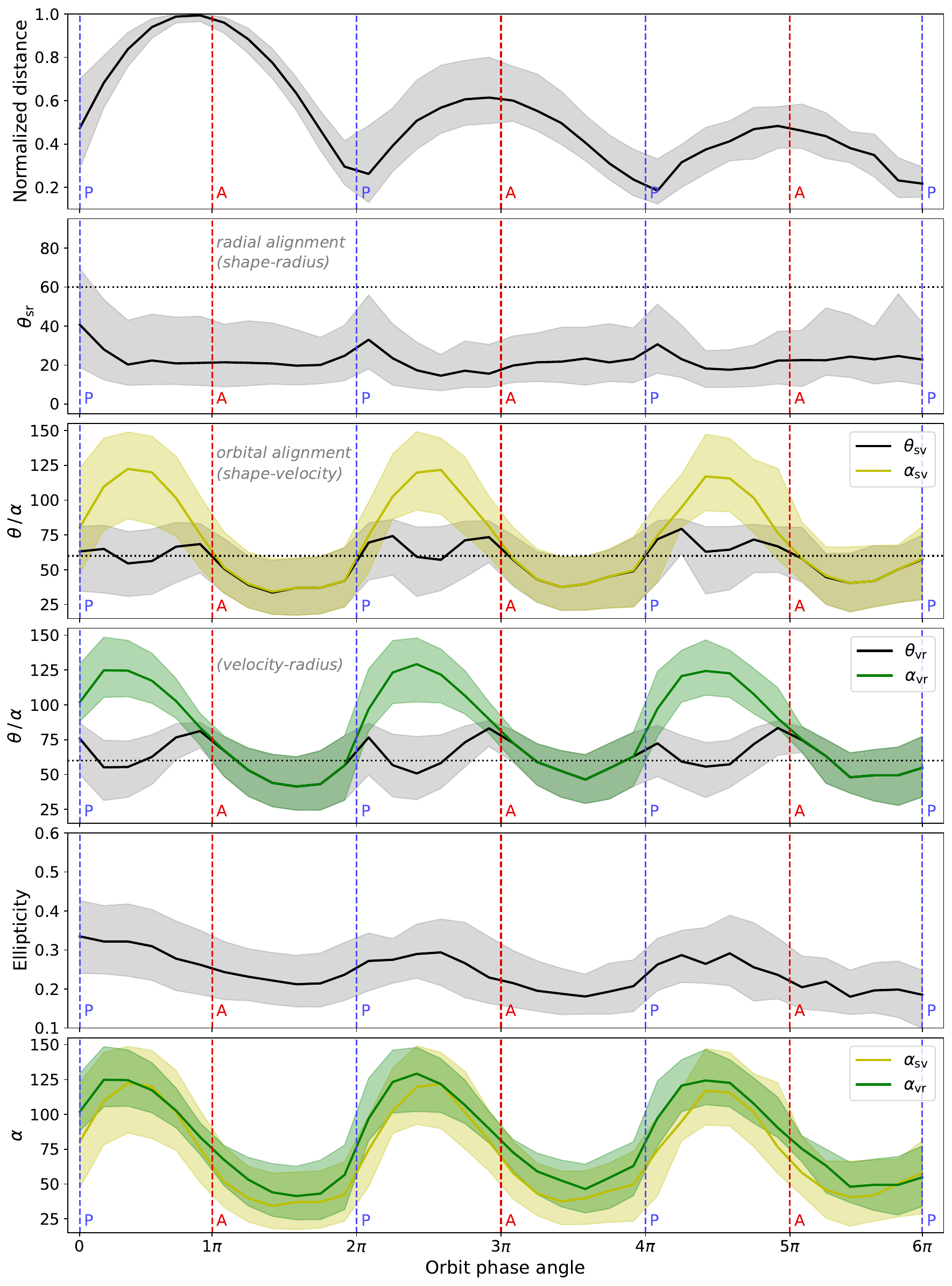}
    \caption{The evolution of different properties along the orbits of subhalos in a host halo mass bin. In all panels, the $x$-axis is the orbit phase angle, which starts from subhalos' first pericentric passages (i.e. orbit phase angle $= 0$) and lasts for the following three periods (i.e. orbit phase angle $= 6\pi$). From top to bottom, these panels show the evolution of the galactocentric distance normalized by its largest value in the orbit, the radial alignment angle $\theta_\mathrm{sr}$ (or $\alpha_\mathrm{sr}$), the orbital alignment angle $\theta_\mathrm{sv}$ (or $\alpha_\mathrm{sr}$), the angle between the galactocentric radial direction and the motion direction $\theta_\mathrm{vr}$, the ellipticity of the subhalos and a direct comparison between $\alpha_\mathrm{sv}$ and $\alpha_\mathrm{vr}$. Note that our original definition of alignment angle (Equation~\ref{eqn:theta}), $\theta$, is between 0 and 90$^\circ$, which is shown as black curves in the third and fourth panels from the top. The definition of $\alpha$ in Equation~\ref{eqn:alpha} ranges from 0 to 180$^\circ$, which is shown as other shallow colored curves (yellow and green, see the legend). The solid curves represent the median values, with shaded regions denoting the 16\% to 84\% percentiles. The blue and red vertical dashed lines represent pericentric (P) and apocentric (A) passages, respectively. The black horizontal dotted lines in the second, third, and fourth panels denote the median value of the angle $\theta$ with random orientations in 3-dimensional space (i.e., $\cos \theta =1/2$).}
    \label{fig:orbit_stack}
\end{figure}

To explore the reasons behind the orbital alignment and its dependence on the satellite/subhalo orbital phase, we trace the evolution of different properties of subhalos along their orbits, since the signals are more prominent in subhalos. We stack complete orbital periods of the subhalos in a host halo mass bin in terms of their orbit phase angles\footnote{The orbit phase angle is defined as $2\pi$ multiplying the ratio between the time elapsed since the pericentric passage $\Delta t$ and the orbital period $T$: $\phi=2\pi \times \,\Delta t/T$ \citep[see e.g., Equation (10) in][]{Han2016}. In this way, $\phi = 0, 2\pi, 4\pi, \dots$ at the pericenter, while $\phi = \pi, 3\pi, 5\pi, \dots$ at the apocenter.} and show their orbital histories for the first three periods in Figure~\ref{fig:orbit_stack}. To explicitly illustrate the variation of alignment angles, we show $\theta_\mathrm{sr}$ and $\theta_\mathrm{sv}$ instead of $\cos{(\theta_\mathrm{sr})}$ and $\cos{(\theta_\mathrm{sv})}$ in this figure. After the subhalo fell into its host halo, the radial alignment angle $\theta_\mathrm{sr}$, remains small ($\sim 20^\circ$) with little variation along the orbit, only slightly increasing near pericenters (black curve in the second panel from the top of Figure~\ref{fig:orbit_stack}). The small variation of $\theta_\mathrm{sr}$ along the orbit is consistent with the relatively weaker dependence of the radial alignment on the orbital phase in Figure \ref{fig:thetasr}. The slight increase of $\theta_\mathrm{sr}$ near pericenters is likely because the subhalo is moving too fast for tidal torque to be completely effective \citep{Pereira2008}.

However, the orbital alignment angle, $\theta_\mathrm{sv}$, exhibits a nearly periodic variation, which is similar to the trend reported in \cite{Pereira2008} (black curve in the third panel from the top of Figure~\ref{fig:orbit_stack}). It shows a decrease during the middle orbital phase both from the pericenter to the apocenter, and from the apocenter to the pericenter, with the latter much more significant. The periodic variation of $\theta_\mathrm{sv}$ results in larger average values of $\theta_\mathrm{sv}$ for subhalos moving outwards, and lower average values of $\theta_\mathrm{sv}$ for those moving inwards, causing the distinct signals of different orbital phases in Figure \ref{fig:thetasv} above. Moreover, when moving from the apocenter to the pericenter, the subhalo features slightly smaller ellipticities.

\begin{figure}[htbp!]
    \centering
    \begin{minipage}{\linewidth}
        \centering
        \includegraphics[width=0.9\textwidth]{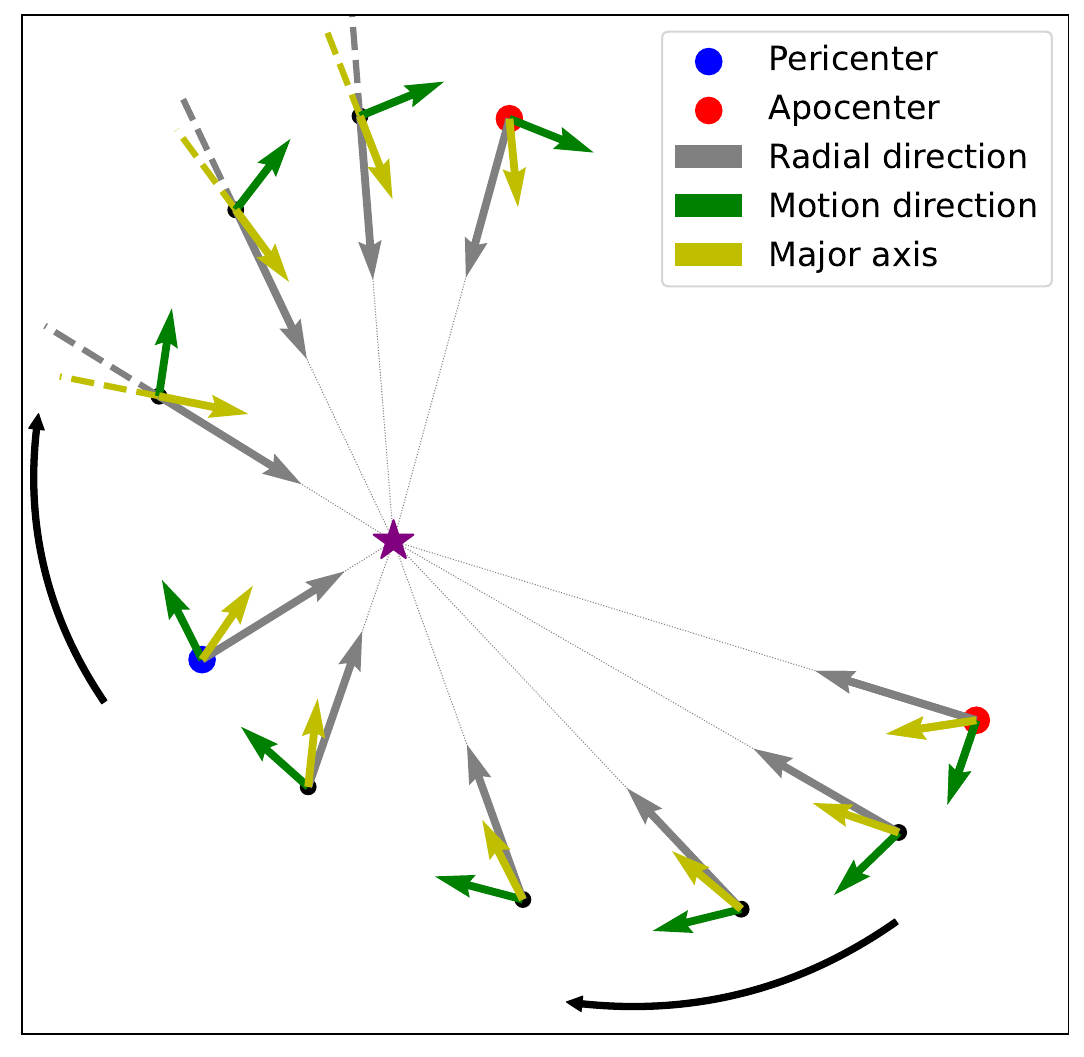}
    \end{minipage}
    \vfill
    \begin{minipage}{\linewidth}
        \centering
        \includegraphics[width=0.9\textwidth]{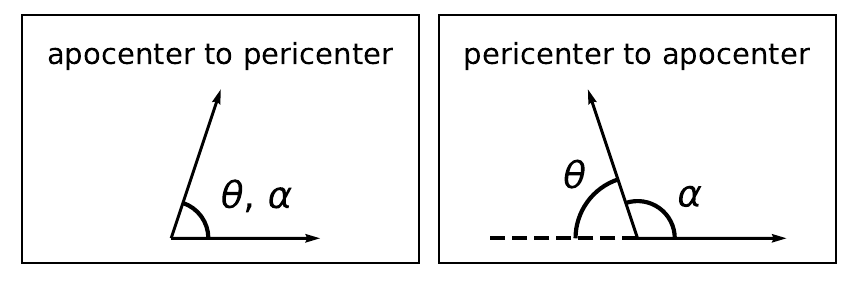}
    \end{minipage}
    \caption{{\bf Top:} The illustration of a representative subhalo orbit around its host in an orbital period. The purple star represents the center of the host. The long black arrows denote the direction of the orbit. Each point represents the location of the subhalo at different time, with the blue and red points denoting the pericenter and the apocenter, respectively. The gray, green and yellow arrows represent the galactocentric radial direction, the motion direction and the major axis of the subhalo at each point, respectively. The points with dashed lines extending arrows backwards denote locations where the angles $\theta$ and $\alpha$ of two vectors differ. {\bf Bottom:} The illustration of the angles $\theta$ (Equation \ref{eqn:theta}) and $\alpha$ (Equation \ref{eqn:alpha}) of two vectors. In the left panel, when the dot product of two vectors is positive, $\theta=\alpha$. While in the right panel, when the dot product of two vectors is negative, $\theta=180^\circ-\alpha$.}
    \label{fig:orbit_ill}
\end{figure}

The periodic variation of $\theta_\mathrm{sv}$ along the orbit may result from the following reasons. First of all, as discussed in Section \ref{subsec: 3d radial alignment}, subhalos' major axes tend to align with their radial directions due to tidal torque throughout its orbits. The orbital alignment we see could be simply caused by the radial alignment. In the top panel of Figure \ref{fig:orbit_ill}, we show a typical orbit for a subhalo from the  apocenter (red dot in the bottom right) to the pericenter (blue dot), and then to the apocenter (red dot on the top) again. Here the latter half of the orbit has decayed in radius to the center (purple star) due to dynamical friction. The angle between the major axes of the subhalo (yellow arrows) and its radial directions pointing to the center (gray arrows), generally well align with each other with small offsets. Near the pericenter, the offset gets slightly larger due to the high velocity, and is consistent with the second panel of Figure~\ref{fig:orbit_stack}. When this subhalo first moves from the apocenter to the pericenter, its orbit is more radial (elongated), and thus its motion directions (green arrows) coincide more with its radial directions, which causes stronger orbital alignment when it moves inwards \citep{Pereira2008}. However, as it approaches the pericenter, the radial direction and the orbit direction becomes more perpendicular, causing increases in $\theta_\mathrm{sv}$. After it passes the pericenter, the orbit decays to smaller radius with respect to the central galaxy and also becomes more circular. As the motion direction of the subhalo gradually reverses, the angle between its motion direction and the major axis (pointing to the central galaxy) becomes greater than 90~$^\circ$ during this latter half of the orbital period. Because of our definition of the orbital alignment angle $\theta_\mathrm{sv}$ (Equation~\ref{eqn:theta}), the orbital alignments we calculated and discussed so far are in fact the angle between the green arrow and the yellow dashed line (pointing away from the central galaxy) during this orbital phase. Nevertheless, due to the more circular orbit at this later phase after decaying, the decrease of $\theta_\mathrm{sv}$ becomes smaller during this latter orbital phase when moving outwards. This generally explained the trend for the black curve in the third panel of Figure~\ref{fig:orbit_stack}. Note if the orbit does not decay and is symmetric from the apocenter to the pericenter, and from the pericenter back to the apocenter, we expect the black curve in the third panel of Figure~\ref{fig:orbit_stack} should be symmetric as well. 

If we define the positive direction for the major axes of subhalos as pointing towards their central galaxies, and allow the orbital alignment angles to range from 0 to 180$^\circ$, the corresponding alignment angle, $\alpha_\mathrm{sv}$ (Equation~\ref{eqn:alpha}) is shown as the yellow curve in the third panel of Figure~\ref{fig:orbit_stack}. It coincides with the $\theta_\mathrm{sv}$ (Equation~\ref{eqn:theta}) definition from the apocenter to the pericenter (see the bottom left panel of Figure~\ref{fig:orbit_ill}), but $\theta_\mathrm{sv}=180^\circ-\alpha_\mathrm{sv}$ from the pericenter to the apocenter (see the bottom right panel of Figure~\ref{fig:orbit_ill}). Because $\alpha_\mathrm{sv}$ show an approximately sine function shape, $\theta_\mathrm{sv}=180^\circ-\alpha_\mathrm{sv}$ then naturally shows a dip in the middle orbital phase from the pericenter to the apocenter.

In the fourth panel of Figure~\ref{fig:orbit_stack}, we also show how the angles between the galactocentric radial direction and the motion direction of subhalos, $\theta_\mathrm{vr}$ and $\alpha_\mathrm{vr}$, change with the orbit phase angles. The trend is largely similar to those in $\theta_\mathrm{sv}$ of the third panel above, and thus directly supports our argument that the orbital alignment is caused by the radial alignment. Note the trends that we discussed in the third and fourth panels are universal among the three orbit periods we investigate in Figure~\ref{fig:orbit_stack}, indicating an orbit like the one in Figure~\ref{fig:orbit_ill} is representative. The key to understand the behaviour of the orbital alignment signal in our study is the orbit decay, so the latter half of the orbit has less radial and orbital alignments. Given the existence of dynamical friction in the simulation, it is natural to expect such orbit decay to universally exist, hence causing the patterns we see in Figure~\ref{fig:orbit_stack}. On the first half orbit from the apocenter to the pericenter, the radius decreases, and dynamical frictions would make the decrease more drastic. Thus the orbit is more radial and elongated. On the latter half orbit from the pericenter back to the apocenter, the radius increases, and dynamical friction would counter-act the increase, making the orbit closer to be circular.

Notably, the green curve ($\alpha_\mathrm{vr}$) in the fourth panel of Figure~\ref{fig:orbit_stack} is overall slightly higher than the yellow curve ($\alpha_\mathrm{sv}$) in the third panel from the top. This is due to the slight offset between the major axes direction and the radial direction, which we have seen from Figure~\ref{fig:orbit_ill} (yellow and gray arrows). This can also be seen from the bottom panel of Figure~\ref{fig:orbit_stack}, where we show a direct comparison between the orbital alignment ($\alpha_\mathrm{sv}$) and the alignment between the galactocentric radial direction and the motion direction of subhalos ($\alpha_\mathrm{vr}$), with the latter showing slightly higher values. This offset results in a gentler decrease from the pericenter to the apocenter of the black curve in the third panel than the fourth one. Note that this offset angle is always towards the motion direction of subhalos (see Figure~\ref{fig:orbit_ill}, where yellow arrows are always biased towards green arrows). We will return to this point slightly later.

We now return to discussing the halo mass dependence of the orbital alignment in Figure~\ref{fig:thetasv}. The stronger orbital alignment signal in more massive halos is likely because satellites/subhalos in more massive host halos have more radial orbits \citep[e.g.][]{Li2020,He2024}. For more radial orbits, the motion directions are more likely to be aligned with the galactocentric radial directions. Therefore, satellites/subhalos in more massive host halos feature stronger orbital alignment signals when moving inwards (blue curves in Figure \ref{fig:thetasv}), while the misalignment signals when moving outwards are weakened (red curves in Figure \ref{fig:thetasv}).

A second reason to explain our measured orbital alignment signals is likely related to tidal stripping. We have mentioned that there is always a small offset between the subhalo's major axis (yellow arrows) and the radial direction (gray arrows) in Figure~\ref{fig:orbit_ill}. The major axes are always slightly biased towards the motion direction (green arrows). This is likely due to tidal streams, which stretch the subhalo along the orbit trajectory. Explicitly, tides cause particles to be stripped from the two Lagrangian points. When they are first stripped, these particles tend to distribute along the radial direction connecting the satellite/subhalo and the center of the host. Afterwards, these stripped particles would move along the leading and trailing arms of the streams formed by tides, as particles moving at the inner/outer sides would move faster/slower with respect to the progenitor, causing elongations along the orbit. Though only bound particles are used in our analysis, and the majority of particles along the leading/trailing arms of the streams are no longer bound to their progenitors, there can still be some particles that have just been stripped and remain loosely bound in the outskirts of their progenitor satellites/subhalos. As these particles get ahead of/fall behind the bulk along the motion direction, they will bias the major axis of the system to the motion direction, leading to the offset between the major axis and the radial direction. 

When the subhalo moves from the apocenter to the pericenter, tidal forces gradually become stronger. More and more particles contribute to this offset between the subhalo's major axis and the radial direction. However, as the subhalo moves around the pericenter, these loosely bound particles in the outskirts might be stripped off, and no longer bound to the subhalo. During the orbital phase to the apocenter after the subhalo passes the pericenter, tidal forces gradually become weaker. The subhalo retains primarily more bound particles, with less particles stripped to form the streams, so the offset becomes weaker, but it still exists (see the bottom panel of Figure~\ref{fig:orbit_stack}). Besides, subhalos moving towards pericenters have slightly rounder shapes with smaller ellipticities (the bottom panel of Figure~\ref{fig:orbit_stack}), probably because the offset is larger and hence subhalos are elongated along two directions (i.e., galactocentric radial directions and motion directions).When moving towards apocenters, they feature slightly larger ellipticities due to the smaller offset and the major elongation along radial directions of their more bound particles.

\subsection{Alignments in projection}
\label{subsec: 2d alignment}

To compare with the alignment signals between the major axes and the motion directions found in MW satellites \citep{Pace2022}, we study the observed signals in projection using the MW/M31-like sample mentioned in Section \ref{sec:data}. For each MW/M31-like central galaxy, we first define its disk plane as the plane perpendicular to the minor axis of the bound star particles inside 20 kpc from the Galactic Center. The observer is then placed at a Galactocentric distance of 8 kpc with a random azimuthal angle\footnote{We have also tried other azimuthal angles, but the results remain nearly the same.} within the disk plane. Each satellite/subhalo is projected on the plane perpendicular to the line-of-sight direction to the observer to simulate the observations of MW satellites. The direction of the projected major axis for each system is computed based on the projected positions of its bound particles onto this projected plane. Note that we use \textit{observed} to denote quantities defined on the 2-dimensional projected plane.

\begin{figure}[htbp!]
    \centering
    \includegraphics[width=0.49\textwidth]{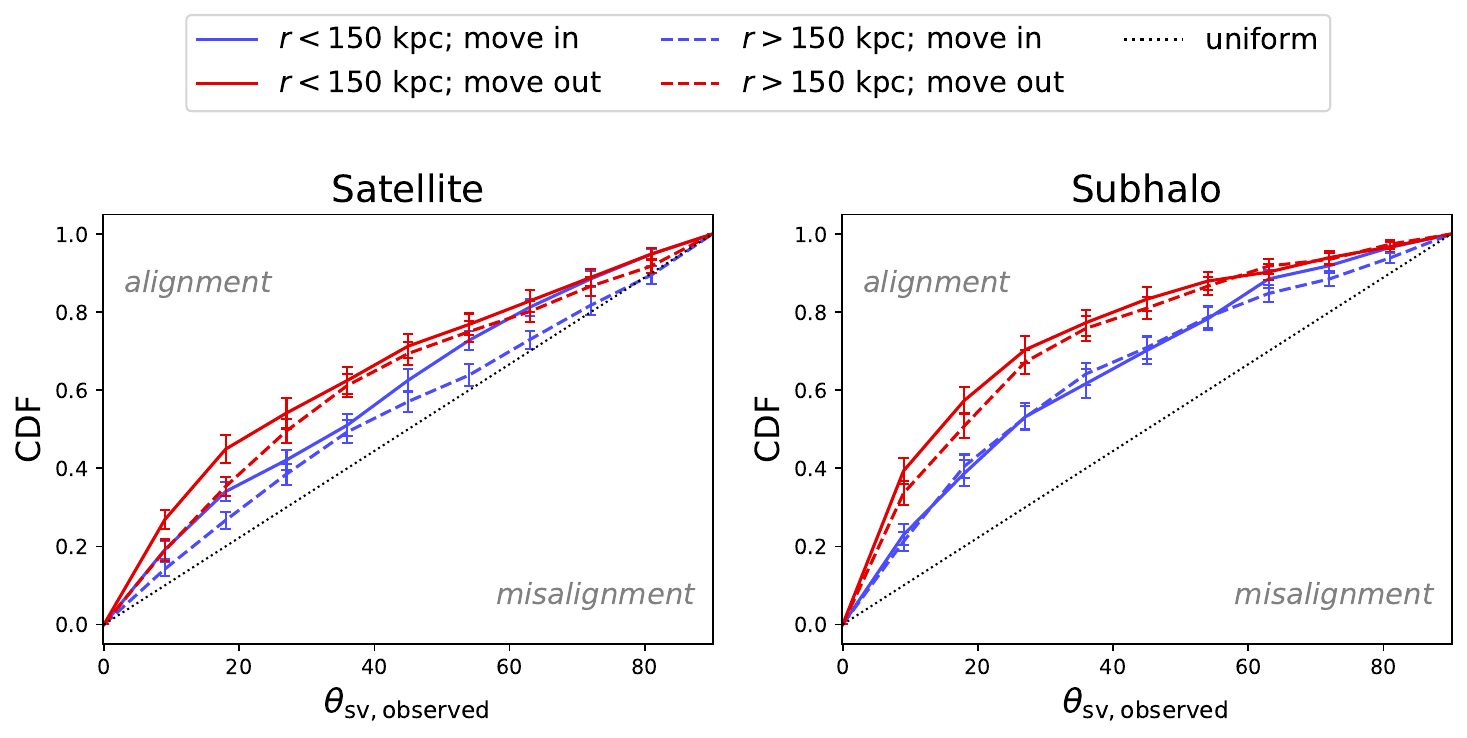}
    \caption{The cumulative distribution function (CDF) of the observed orbital alignment angle $\theta_\mathrm{sv,observed}$ between the projected major axes of the systems and their projected motion directions, for satellites/subhalos surrounding MW/M31-like central galaxies. The left column shows the results of satellites, and the right column shows the results of subhalos. The blue (red) curves refer to systems moving inwards (outwards). The solid (dashed) curves refer to systems at the galactocentric radii $r$ smaller (larger) than 150kpc. The black dotted line in each panel denotes the CDF of a uniform distribution, and curves above (below) it correspond to alignment (misalignment) signals. The error bars represent the 1$\sigma$ scatters of 50 bootstrapped subsamples of the systems in the same bin.}
    \label{fig:thetasv_obs}
\end{figure}

Figure~\ref{fig:thetasv_obs} presents the observed orbital alignment angle between the projected major axes of satellites/subhalos in the MW/M31-like sample and their projected motion directions. Note in 2-dimensional space, it is the orbital alignment angle $\theta_\mathrm{sv,observed}$ itself that follows a uniform distribution for random orientations (not $\cos(\theta_\mathrm{sv,observed})$), with the uniform distribution denoted by the diagonal black dotted line in each panel. Here, in contrast to 3-dimensional cases in Figure \ref{fig:thetasr} and \ref{fig:thetasv}, curves above the diagonal black lines mean alignment. Similar to the situation in 3-dimensional space, the alignment signals of subhalos are also more prominent than those of satellites in this 2-dimensional projected space. However, for satellites/subhalos moving outwards, they feature even stronger alignment signals than those moving inwards, in contrast to the misalignment signals in previous Figure~\ref{fig:thetasv}.

\begin{figure}[htbp!]
    \centering
    \includegraphics[width=0.49\textwidth]{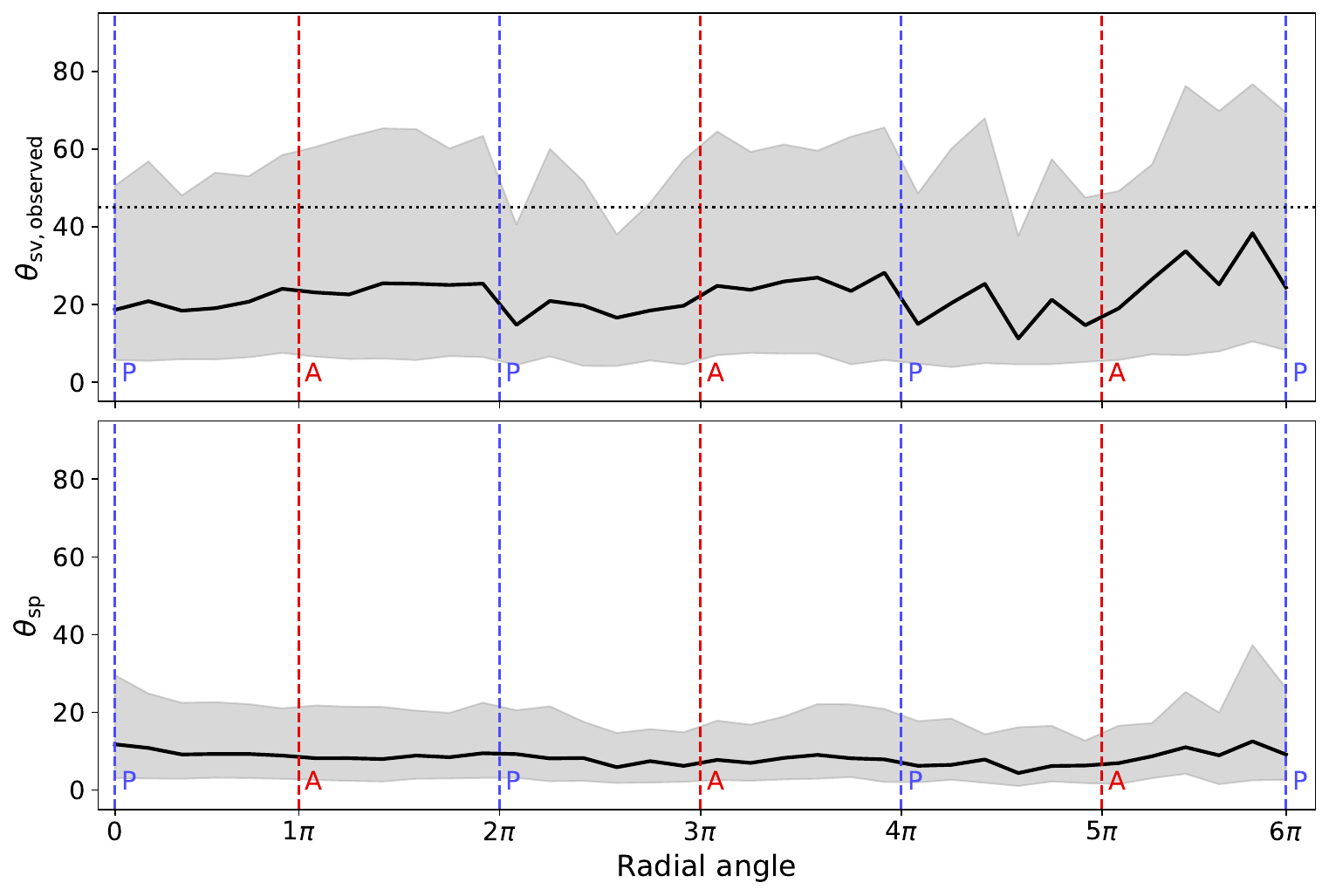}
    \caption{Similar to Figure \ref{fig:orbit_stack}, but of the observed orbital alignment angle $\theta_\mathrm{sv,observed}$ of subhalos in our MW/M31-like sample (top panel) and the angle $\theta_\mathrm{sp}$ (bottom panel), which is the angle between the major axes of the subhalos and their orbital planes. The black horizontal dotted line in the top panel denotes the median value of the angle $\theta$ with random orientations in 2-dimensional space (i.e., $\theta =45^\circ$).}
    \label{fig:orbit_stack_ob}
\end{figure}

Similar to Section \ref{subsec:cause}, we trace the evolution of the observed orbital alignment angle $\theta_\mathrm{sv,observed}$ of subhalos, as shown in the top panel of Figure \ref{fig:orbit_stack_ob}. In spite of large uncertainties (shaded regions), $\theta_\mathrm{sv,observed}$ remains below $45^\circ$ along the orbit, leading to alignment signals for subhalos both moving inwards and outwards and showing almost no dependence on orbit phase angle. Moreover, uncertainties of $\theta_\mathrm{sv,observed}$ tend to be sightly smaller for systems moving outwards (from pericenters to apocenters). Different from the alignment in 3-dimensional space, the alignment in projection may be mainly due to projection effects. In a stable configuration, the major axis of a satellite/subhalo tends to lie within the orbital plane, which is defined by its motion direction and galactocentric radial direction, otherwise, the satellite/subhalo has to spin with a frequency equalling to or integer times of the orbital frequency. We illustrate the evolution of the angle between the major axes of subhalos and their orbital planes, which is referred to as $\theta_\mathrm{sp}$, in the bottom panel of Figure \ref{fig:orbit_stack_ob}. As expected, $\theta_\mathrm{sp}$ remains small ($\sim 10^\circ$) with little uncertainties along the orbit. Moreover, located near the Galactic Center, the observer is also approximately coplanar with these directions\footnote{This argument requires that the Galactocentric distance of satellites/subhalos are much larger than that of the observer. We show the Galactocentric distance distribution of satellites/subhalos in Appendix \ref{appsec:radial distribution} to verify it.}. Thus, the projected major axis of the system will always align with its projected motion direction.

\begin{figure}[htbp!]
    \centering
    \begin{minipage}{\linewidth}
        \centering
        \includegraphics[width=0.98\linewidth]{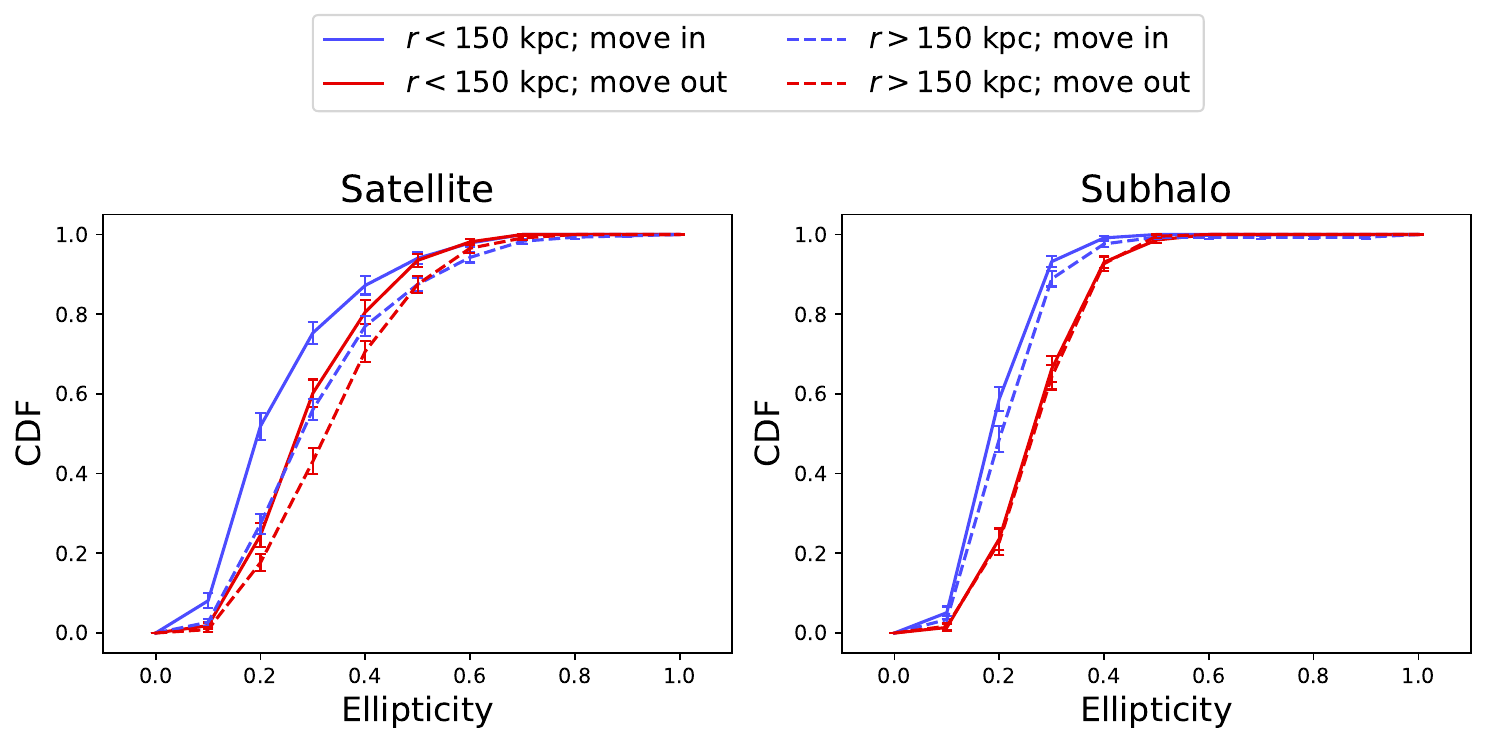}
    \end{minipage}
    \vfill
    \begin{minipage}{\linewidth}
        \centering
        \includegraphics[width=0.98\linewidth]{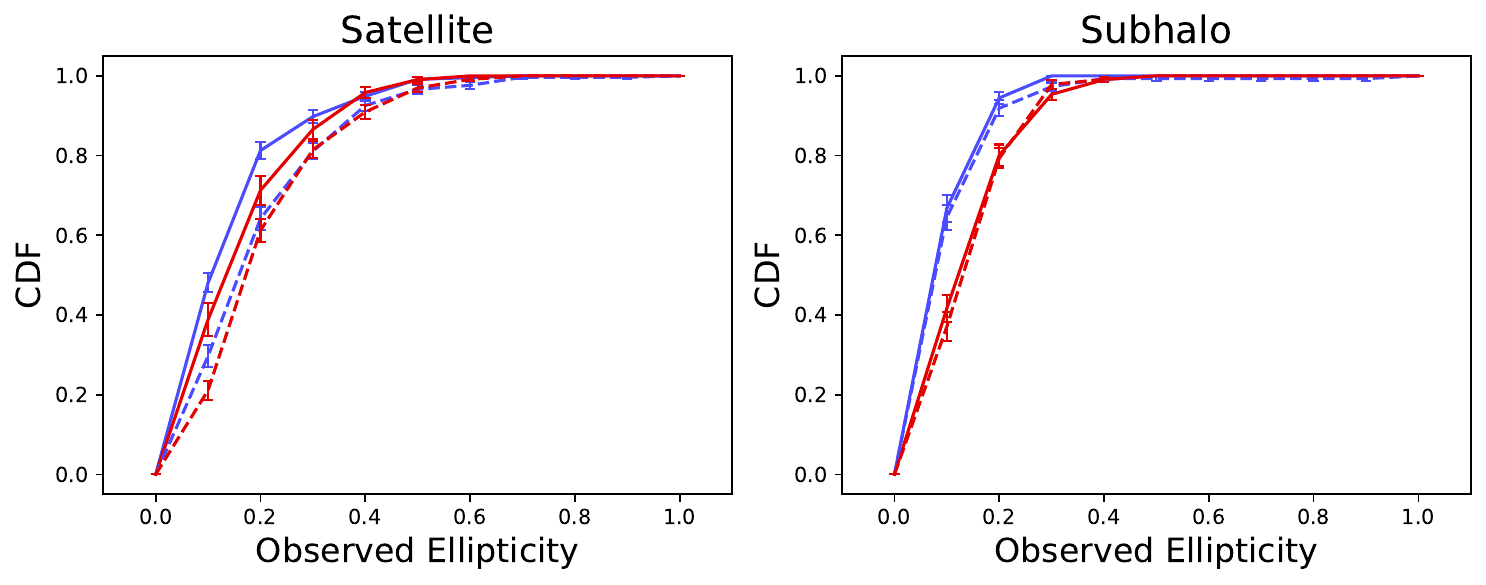}
    \end{minipage}
    \caption{Similar to Figure \ref{fig:thetasv_obs}, but shows the cumulative ellipticity (top) and observed ellipticity (bottom) distribution of satellites/subhalos. The ellipticities are calculated from the spatial distributions of bound star/dark matter particles for satellites/subhalos in 3-dimensional space, while the observed ellipticities are calculated from those on the projected plane perpendicular to the mock line of sight.}
    \label{fig:ell}
\end{figure}

Due to relatively round shapes of satellites/subhalos, the projected major axes computed from their bound particles may slightly deviate from the orbital plane, which results in imperfect alignment signals ($\theta_\mathrm{sv,observed}<45^\circ $ but $\theta_\mathrm{sv,observed}>0^\circ $). Orientations of more elliptical systems will be better determined, and these systems will have more noticeable alignment signals. As mentioned in Section \ref{subsec:cause}, satellites/subhalos heading towards apocenters tend to be more elongated. We show CDFs of the ellipticities and the observed ellipticities of satellites/subhalos in Figure \ref{fig:ell}. Indeed, satellites/subhalos moving outwards have larger ellipticities and hence larger observed ellipticities, and therefore are more likely to be observed with stronger orbital alignment signals in projection.

\begin{figure}[htbp!]
    \centering
    \includegraphics[width=0.45\textwidth]{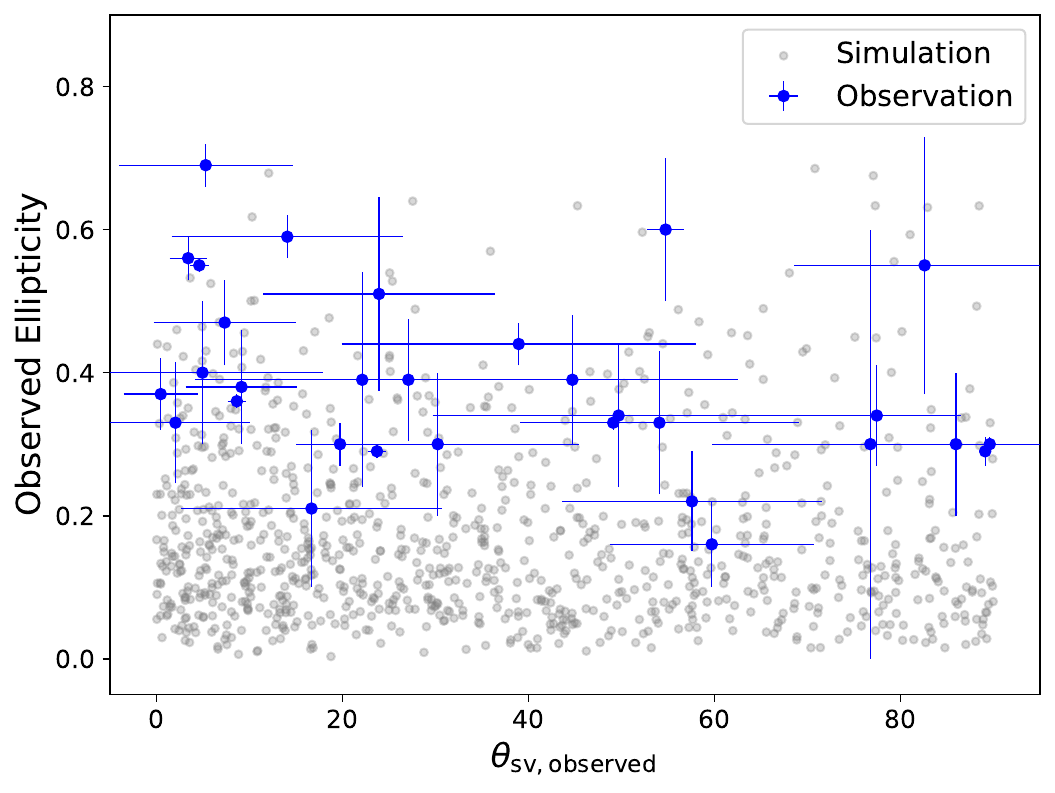}
    \caption{The observed orbital alignment angle $\theta_\mathrm{sv,observed}$ of satellites versus their observed ellipticity. The grey dots denote satellites in the TNG50 MW/M31-like sample, and blue dots with error bars denote MW satellites in \cite{Pace2022}.}
    \label{fig:simu_mw_comp}
\end{figure}

In Figure \ref{fig:simu_mw_comp}, we compare our results for satellites in the TNG50 MW/M31-like sample with those for MW satellites in \cite{Pace2022}. Generally, the trends are the same in both the simulation and observation. Satellites with larger observed ellipticities feature smaller alignment angles $\theta_\mathrm{sv,observed}$, or namely more prominent alignment signals. However, the observed ellipticities of MW satellites (blue dots with error bars) are systematically larger than those of satellites in TNG50 MW/M31-like sample (grey dots). The larger observed ellipticities for MW satellites can be caused by several reasons. First, the observed ellipticity of each MW satellite in \cite{Pace2022} is based on a projected Plummer model \citep{Plummer1911}, which includes prior assumptions that might cause some bias. On the other hand, we only use satellites in the TNG50 MW/M31-like sample that have at least 100 star particles for reliable measurements on their shapes. Due to the resolution of the simulation, this results in the selection of simulated satellites more massive than most of MW satellites in \cite{Pace2022}, which may also bias our sample. Moreover, it has been pointed out in \cite{Alonso2023} that satellite disruptions are stronger in TNG than real observations, which could also influence our shape measurements and affect fair comparisons. Lastly, in real observation only the brighter central parts of each satellite can be observed, and sparse member stars in outskirts of each satellite might have been missed. However, in simulation we include all physically bound particles, which may complicate the exact comparisons.

Constructing a mock observation for simulated TNG50 MW/M31-like sample by considering all observational factors into account is, however, beyond the scope of this paper. Despite the difference in the measured ellipticities, the general trends are consistent between observed MW satellites and simulated satellites in our TNG50 MW/M31-like sample. We believe our main interpretations are valid.

\section{Conclusions}
\label{sec:concl}

In this paper, we use the IllustrisTNG50 simulation to study two different alignment signals of satellites/subhalos. The first is the alignment between the major axes of satellites/subhalos and their galactocentric radial directions (radial alignment), and the second is the alignment between the major axes of satellites/subhalos and their motion directions (orbital alignment).

We find significant radial alignment signals for both satellites and their subhalos, with those for subhalos more prominent. Tidal torque by host halos is the main cause of the radial alignment. The stronger alignment signals of subhalos are due to the fact that subhalos are more extended and can be more easily influenced by tides. The strength of the radial alignment shows almost no dependence on the host halo mass, which indicates that gravitational forces are self-similar or scale free. On the other hand, satellites in more central regions of their hosts show stronger signals than those in outer regions due to the stronger tides in the center. Besides, satellites moving outwards feature stronger radial alignment signals than those moving inwards, whereas subhalos moving inwards feature stronger signals. This is likely because subhalos are more extended, and can respond to tidal torque more quickly. Satellites embedded in subhalos need longer time to be influenced, and thus their radial alignments are stronger after passing pericenters and moving outwards.

The orbital alignment is weaker than the radial alignment, but is significant. Similar to radial alignments, subhalos exhibit more prominent orbital alignment signals than satellites, and satellites/subhalos in inner regions feature relatively stronger signals than those in outer regions. Interestingly, we find that the orbital alignment signals are opposite for systems moving inwards and outwards. Satellites/subhalos moving inwards tend to have their major axes aligned with their motion directions, leading to orbital alignment signals. In contrast, those moving outwards tend to have their major axes perpendicular to their motion directions, leading to misalignment signals. With the increase of the host halo mass, the alignment signals of satellites/subhalos moving inwards become stronger, whereas the misalignment signals of satellites/subhalos moving outwards become weaker.  

To explore the mechanism of the orbital alignment, we trace the evolution of different properties of subhalos along their orbits as in \cite{Pereira2008}. The orbital alignment angle $\theta_\mathrm{sv}$, which is defined as the angle between the major axes of satellites/subhalos and their motion directions, presents a periodic variation between two pericenters. The lower average values of $\theta_\mathrm{sv}$ for subhalos moving inwards correspond to alignment signals, and larger average values of $\theta_\mathrm{sv}$ for subhalos moving outwards correspond to misalignment signals. Besides, the ellipticities of subhalos are slightly smaller when they move inwards. 

There are two possible reasons for the periodic variation of $\theta_\mathrm{sv}$. Firstly, the orbits from apocenters to pericenters are more radial due to dynamical frictions, the motion directions of subhalos would coincide more with their radial directions, and radial alignments causes orbital alignments. After they pass pericenters to apocenters, the orbits become more circular also due to dynamical frictions, and thus the motion directions of subhalos become more perpendicular to the radial directions, causing misalignment signals. Moreover, the orbital alignment signals become stronger with the increase of the host halo mass, likely because satellites/subhalos in more massive host halos have more radial orbits as reported in previous studies \citep[e.g.][]{Li2020,He2024}. 

Secondly, particles of subhalos that have just been stripped but remain loosely bound in the outskirts of the systems, which is about to form streams, stretch along the orbit trajectories and may contribute to the orbital alignment. As subhalos move from apocenters to pericenters, gradually stronger tidal forces cause stronger stripping, and hence stronger alignment signals. At the pericenter, particles in the outskirts are stripped off due to the strongest tides. When subhalos head for the apocenter, gradually weaker tidal forces strip less particles, and subhalos are mainly directed to their radial directions, misaligned with their motion directions. Additionally, since subhalos moving inwards are elongated along two directions (galactocentric radial directions and motion directions), they feature smaller ellipticities.

In the end and in order to draw connections to the observed orbital alignment signals of our Milky Way (MW) satellites, we study the orbital alignment in the projected plane for the MW/M31-like sample in the TNG50 simulation \cite{Engler2021,Engler2023,Pillepich2024}. Here the projected plane is perpendicular to the line-of-sight direction of the mock observer. \cite{Pace2022} found that, for MW dwarf spheroidal satellite galaxies, the major axes of those with large ellipticities (ellipticities larger than 0.4) tend to be aligned with their proper motion directions. We find that their observed orbital alignment signals are due to the coplanarity of the satellite/subhalo's major axis with the orbital plane, which is defined by the galactocentric radial direction and the motion direction of the satellite/subhalo. Since the observer also approximately lies in the orbital plane, satellites/subhalos' projected major axes are always aligned with their proper motion directions, causing orbital alignment signals in projection. Since satellites/subhalos moving outwards have larger ellipticities, their projected major axes can be more accurately determined and we find that satellites/subhalos moving from pericenters to apocenters feature stronger orbital alignment signals in projection. Finally, we compare our results for satellites in simulation with those for MW satellites in \cite{Pace2022}. Both show the same trend that satellites with larger observed ellipticities have stronger orbital alignment signals in projection, although the observed ellipticities of MW satellites are systematically larger.

Our study in this paper provides interesting discussions on the alignment of satellite galaxies in 3-dimensional space and in projected case analogous to real MW satellite observations \citep{Pace2022}. Unfortunately, the major axes of MW satellites are artificially elongated along the line-of-sight direction due to inaccurate distance measurements \citep[e.g.][]{2025arXiv250322800L}. In the future, if the precision in stellar distance measurements can be significantly improved to enable precise measurements of satellite shapes and orientations based on their member star spatial distributions, it would be promising for us to investigate their radial and orbital alignments in 3-dimensional space.

\section*{Acknowledgments}

We thank Andrew Pace for sharing the data and associated code in his work. This work is supported by NSFC (12573022, 12595312, 12273021), the National Key R\&D Program of China (2023YFA1605600, 2023YFA1605601), 111 project (No.\ B20019), and the Office of Science and Technology, Shanghai Municipal Government (grant Nos. 24DX1400100, ZJ2023-ZD-001). We thank the sponsorship from Yangyang Development Fund. The computations of this work are carried on the Gravity supercomputer at the Department of Astronomy, Shanghai Jiao Tong University. HY is supported by T.D. Lee scholarship. SK acknowledges support from the Science \& Technology Facilities Council (STFC) grant ST/Y001001/1.

For the purpose of open access, the author has applied a Creative
Commons Attribution (CC BY) licence to any Author Accepted
Manuscript version arising from this submission.

\appendix

\section{Alignments in 3-dimensional space for the MW/M31-like sample}
\label{appsec:3d alignment mw sample}

\begin{figure}[htbp!]
    \centering
    \includegraphics[width=0.49\textwidth]{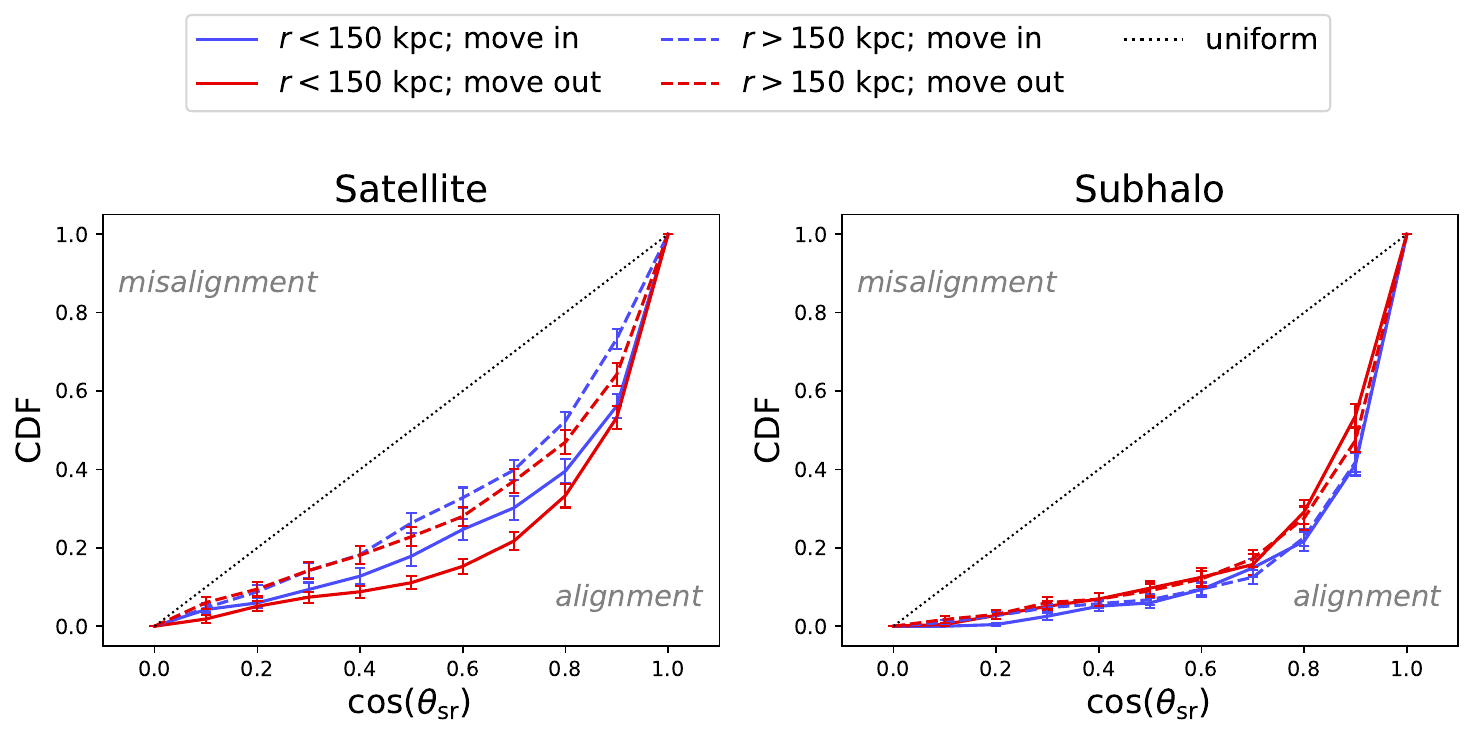}
    \caption{Similar to Figure \ref{fig:thetasr}, but showing the radial alignment signals for satellites/subhalos in the MW/M31-like sample.}
    \label{fig:thetasr_mw}
\end{figure}

\begin{figure}[htbp!]
    \centering
    \includegraphics[width=0.49\textwidth]{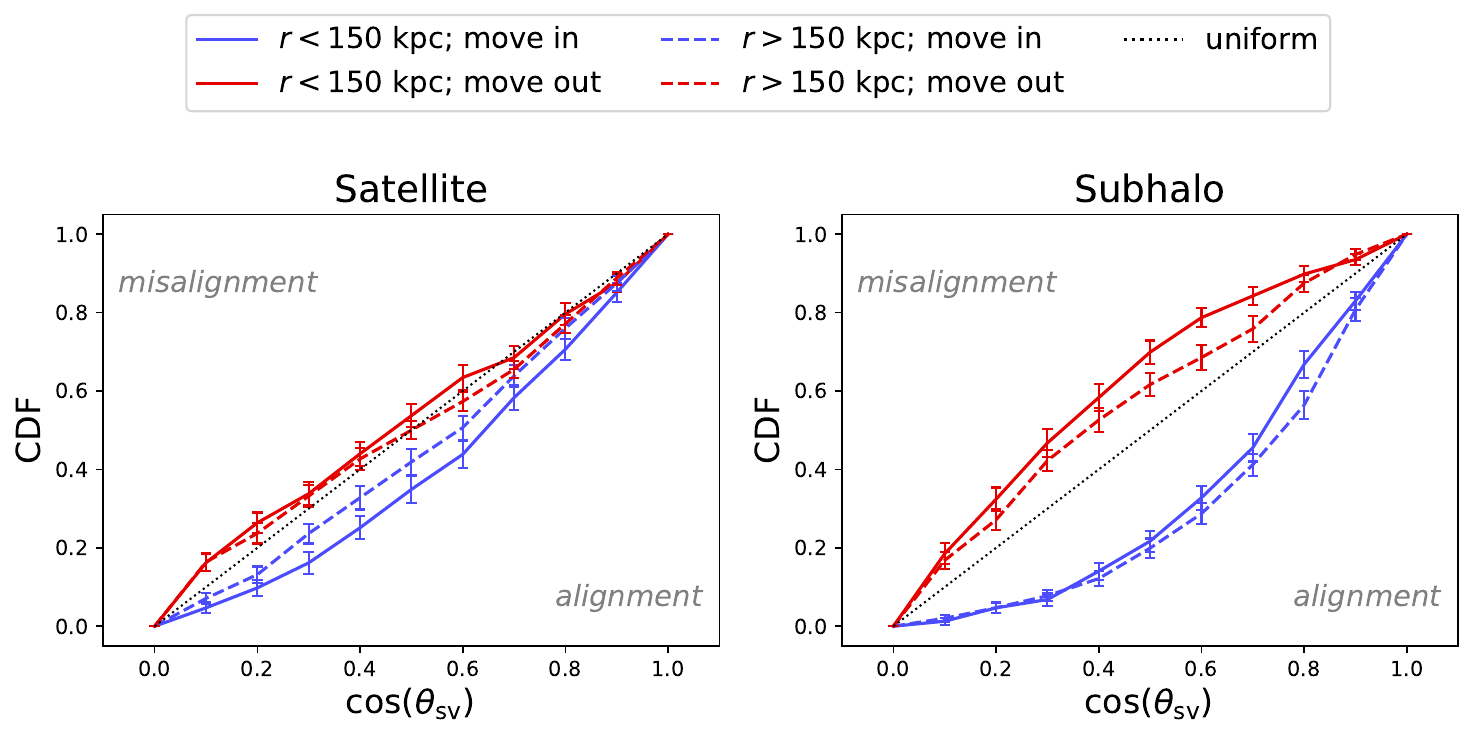}
    \caption{Similar to Figure \ref{fig:thetasv}, but showing the orbital alignment signals for satellites/subhalos in the MW/M31-like sample.}
    \label{fig:thetasv_mw}
\end{figure}

We check the radial and orbital alignments in 3-dimensional space for satellites/subhalos in the MW/M31-like sample, and present results in Figures \ref{fig:thetasr_mw} and \ref{fig:thetasv_mw}, which are consistent with those in Figures \ref{fig:thetasr} and \ref{fig:thetasv} of the main text. The numbers of systems in different bins are provided in Table \ref{tab:num per bin mw}.

\begin{table*}
\centering
\begin{tabular*}{0.8\textwidth}{@{\extracolsep{\fill}}cccccc}
\hline
\multirow{3}{*}{}  & \multicolumn{5}{c}{Number of satellites/subhalos in the MW/M31-like sample}                                                              \\ \cline{2-6} 
                   & \multirow{2}{*}{Total} & $r<\mathrm{150 \,kpc}$ & $r<\mathrm{150 \,kpc}$ & $r>\mathrm{150 \,kpc}$ & $r>\mathrm{150 \,kpc}$ \\
                   &                        & move in             & move out            & move in             & move out            \\ \hline
MW/M31-like sample & 979                    & 235                 & 216                 & 296                 & 232                 \\ \hline
                   &                        &                     &                     &                     &                    
\end{tabular*}
\caption{Similar to Table \ref{tab:num per bin}, but showing numbers of satellites/subhalos in the MW/M31-like sample with different Galactocentric distances and with different directions of radial velocities.}
\label{tab:num per bin mw}
\end{table*}

\section{Results from the reduced MoI tensor}
\label{appsec:red moi}

\begin{figure}[htbp!]
    \centering
    \includegraphics[width=0.49\textwidth]{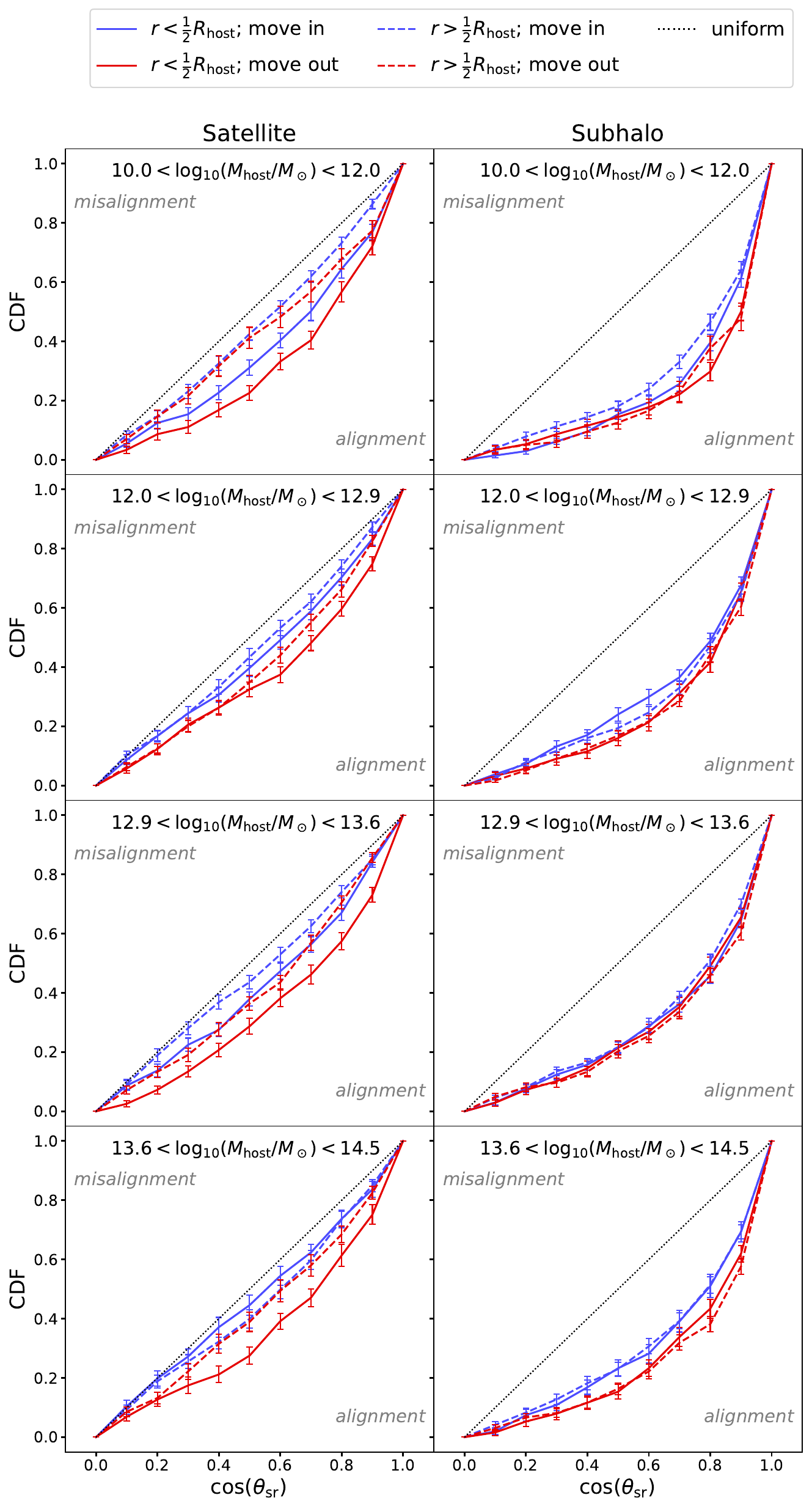}
    \caption{Similar to Figure \ref{fig:thetasr}, but the orientation of a satellite/subhalo is determined by the reduced MoI tensor.}
    \label{fig:thetasr_r}
\end{figure}

\begin{figure}[htbp!]
    \centering
    \includegraphics[width=0.49\textwidth]{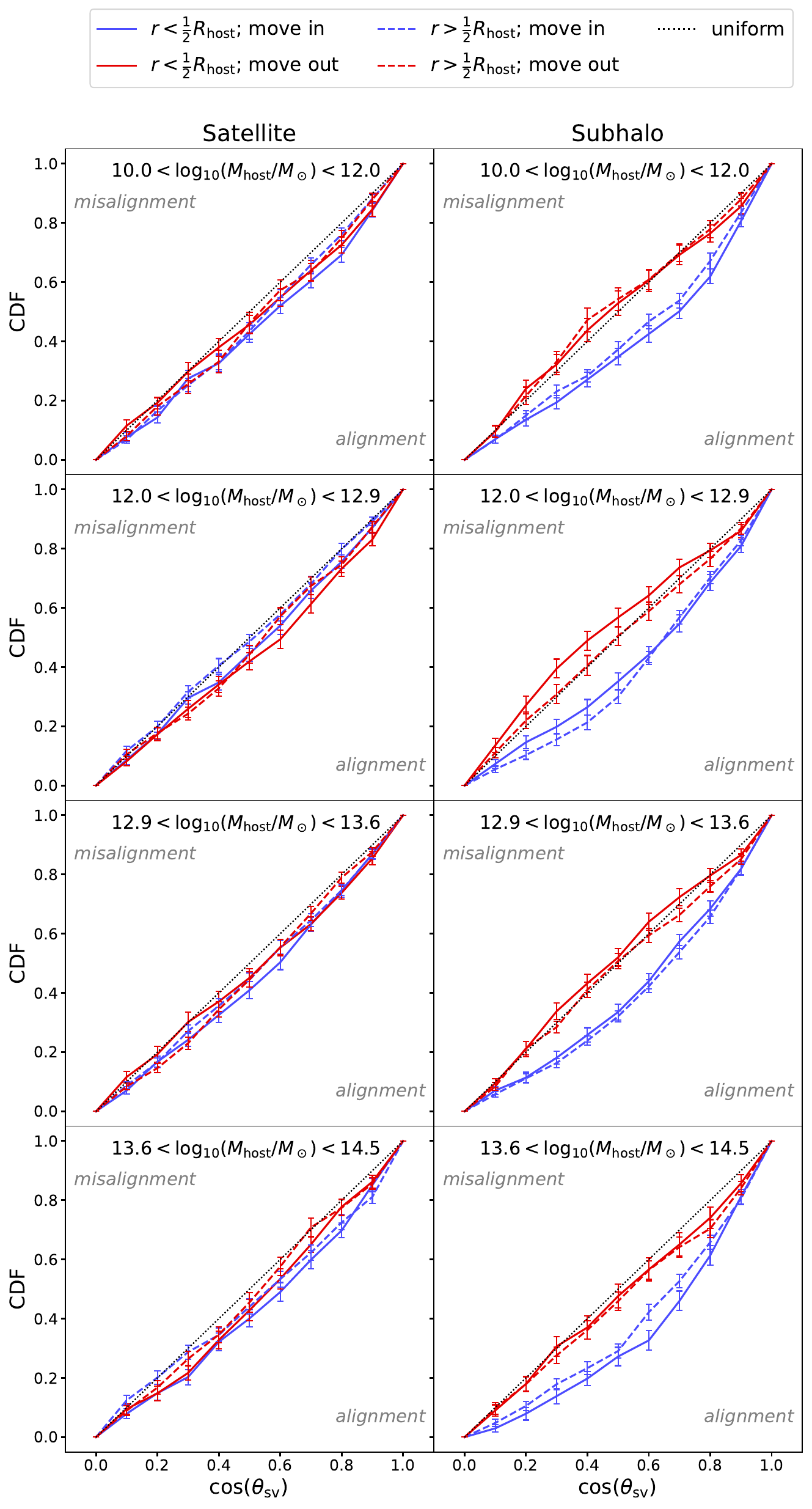}
    \caption{Similar to Figure \ref{fig:thetasv}, but the orientation of a satellite/subhalo is determined by the reduced MoI tensor.}
    \label{fig:thetasv_r}
\end{figure}

Figures~\ref{fig:thetasr_r} and \ref{fig:thetasv_r} show results based on satellite/subhalo orientations computed from the reduced MoI tensor, which have consistent trends with those from the original MoI tensor shown in Section \ref{subsec: 3d radial alignment} and \ref{subsec: 3d orbital alignment} in the main text, but the alignment signals are less significant due to larger weights assigned to particles in more inner regions.

\section{Distance distribution of satellites/subhalos in the MW/M31-like sample}
\label{appsec:radial distribution}

\begin{figure}[htbp!]
    \centering
    \includegraphics[width=0.49\textwidth]{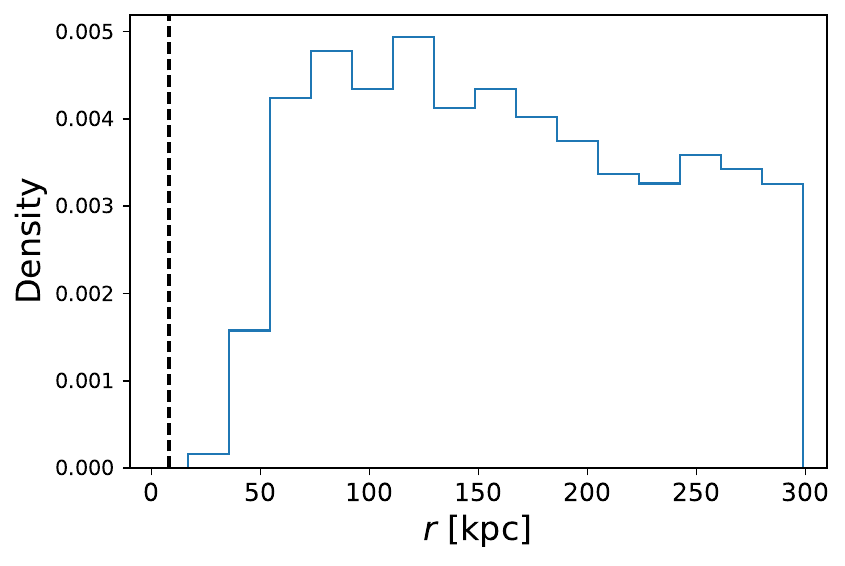}
    \caption{The distribution of Galactocentric distances for satellites/subhalos in the MW/M31-like sample. The black vertical dashed line denotes the location of the observer, placed at a Galactocentric distance of 8 kpc.}
    \label{fig:r distr}
\end{figure}

The orbital alignment in projection is due to the coplanarity of the observer with the orbital planes of satellites/subhalos, which is valid when the Galactocentric distances of satellites/subhalos are much larger than that of the observer (i.e. 8kpc). In Figure \ref{fig:r distr} we show the Galactocentric distance distribution of satellites/subhalos in the MW/M31-like sample. Most satellites/subhalos are much far from the Galactic Center.

\bibliography{master}

\end{document}